\providecommand{\columnwiselinenumbersfalse}{}%
\providecommand{\runningpagewiselinenumbers}{}%
\PassOptionsToPackage{desactivate}{linenoaa}

\documentclass{aa}  
\usepackage{graphicx}
\usepackage{txfonts}
\usepackage{lipsum}
\usepackage{lscape}
\usepackage{placeins}

\usepackage{amsmath}
\usepackage{amssymb}
\usepackage{mathrsfs}
\usepackage{booktabs}
\usepackage{listings}
\usepackage{anyfontsize}

\usepackage[svgnames]{xcolor}
\usepackage{hyperref}
\usepackage{orcidlink}
\let\CheckCommand\providecommand
\usepackage{microtype}

\newcommand{\reddtext}[1]{#1}

\definecolor{C0blue}{RGB}{31,119,180}
\definecolor{C1orange}{RGB}{255,127,14}
\definecolor{C2green}{RGB}{44,160,44}
\definecolor{C3red}{RGB}{214,39,40}
\definecolor{C4purple}{RGB}{148,103,189}
\definecolor{C5brown}{RGB}{140,86,75}
\definecolor{C6pink}{RGB}{227,119,194}
\definecolor{C7gray}{RGB}{127,127,127}
\definecolor{C8olive}{RGB}{188,189,34}
\definecolor{C9cyan}{RGB}{23,190,207}

\hypersetup{
    breaklinks=true,
    colorlinks=true,
    linkcolor=C2green,
    citecolor=C0blue,
    filecolor=magenta,      
    urlcolor=C9cyan,
    pdftitle={reddemcee - Pe{\~n}a, P. and Jenkins, J.},
    pdfauthor={Pe{\~n}a, Pablo},
    }

\newcommand{\reddemc}{\texttt{reddemcee}}
\newcommand{\dynesty}{\texttt{dynesty}}

\newcommand{\lzest}{$\ln\widehat{\mathcal{Z}}$}
\newcommand{\lzerrest}{$\widehat{\sigma}_{\ln{\widehat{\mathcal{Z}}}}$}

\newcommand{\zdiff}{$\Delta_{\mathcal{Z}}$}
\newcommand{\zlogl}{$\mathcal{L}(\widehat{\mathcal{Z}})$}

\newcommand{\Uniform}[2]{$\sim$$\mathcal{U}(#1, #2)$}
\newcommand{\Normal}[2]{$\sim$$\mathcal{N}(#1, #2)$}

\newcommand{\refsec}[1]{\hyperref[#1]{Sect\,\ref*{#1}}}
\newcommand{\refeq}[1]{\hyperref[#1]{Eq.\,\ref*{#1}}}
\newcommand{\reffig}[1]{\hyperref[#1]{Fig.\,\ref*{#1}}}
\newcommand{\reftab}[1]{\hyperref[#1]{Table\,\ref*{#1}}}
\newcommand{\refapp}[1]{\hyperref[#1]{Appendix\,\ref*{#1}}}

\begin{document}
   \title{Closing the evidence gap: reddemcee, a fast adaptive parallel tempering sampler}

   \subtitle{Next-generation ladder adaptation and evidence estimators for parallel tempering}

   \author{Pablo A. Pe{\~n}a R.\orcidlink{0000-0002-8770-4398}\inst{1,2} \and James S. Jenkins\orcidlink{0000-0003-2733-8725}\inst{1,2}}

   \institute{Instituto de Estudios Astrof\'isicos, Facultad de Ingenier\'ia y Ciencias, Universidad Diego Portales, Av. Ej\'ercito 441, Santiago, Chile
            \and Centro de Astrof\'isica y Tecnolog\'ias Afines (CATA), Casilla 36-D, Santiago, Chile}

   \date{Revised version received by journal}

\abstract{Markov chain Monte Carlo (MCMC) excels at sampling complex posteriors, but traditionally lags behind nested sampling in accurate evidence estimation, which is crucial for model comparison in astrophysical problems.}{We introduce \reddemc, an adaptive parallel tempering ensemble sampler, aiming to close this gap by simultaneously presenting next-generation automated temperature-ladder adaptation techniques and robust, low-bias evidence estimators.}{\reddemc~couples an affine-invariant stretch move with five interchangeable ladder-adaptation objectives--uniform swap-acceptance rate, swap mean distance, Gaussian area overlap, small Gaussian gap, and equalised thermodynamic length--implemented through a common differential update rule. Three evidence estimators are provided: curvature-aware thermodynamic integration (TI+), geometric-bridge stepping stones (SS+), and a novel hybrid algorithm that blends both approaches (H+). Performance and accuracy of the sampler are benchmarked on n-dimensional Gaussian shells, Gaussian egg-box, Rosenbrock functions, and the real exoplanet radial-velocity time-series dataset of HD\,20794.}{Across shells up to 15 dimensions, \reddemc~achieves roughly 7 times the effective sampling speed of the best dynamic nested sampling configuration. The TI+, SS+ and H+ estimators recover estimates to within |$\Delta\ln\mathcal{Z}$|$\lesssim$3\% and supply realistic error bars with as few as six temperatures. In the HD\,20794 case study, \reddemc~reproduces literature model rankings and yields tighter yet consistent planetary parameters compared with \dynesty, with evidence errors that track run-to-run dispersion.}{By unifying fast ladder adaptation with reliable evidence estimators, \reddemc~delivers strong  throughput and accurate evidence estimates, often matching, and occasionally surpassing, dynamic nested sampling, while preserving the rich posterior information that makes MCMC indispensable for modern Bayesian inference.}

   \keywords{Methods: numerical -- 
                Methods: statistical -- 
                Techniques: radial velocities --
                Planets and satellites: individual: HD 20794}

   \maketitle

\section{Introduction}  \label{sec:intro}

In scientific research, any measured data needs to be tested \reddtext{against} a suitable hypothesis. As such, robust statistical techniques are indispensable for parameter estimation and model comparison. Science frequently faces the challenge of characterising probability distributions arising from complex, high-dimensional models \reddtext{ridden} with nuisance parameters, uncertainties, and degeneracies.
Markov chain Monte Carlo (MCMC) methods are widely recognised as a vital tool in many areas of science. They enable researchers to characterise parameter uncertainties precisely and compare models rigorously. Research areas that rely on precise inference, such as phylogenetics \citep{phylogenetics1, phylogenetics2}, physio-chemistry \citep{hansmann, sugita99}, gravitational waves \citep{grav_wav1, grav_wav2}, and exoplanet discovery \citep{2020NatAs...4.1148J, jvines-2023-minineptune}, benefit from the flexibility of MCMC techniques to explore difficult posterior landscapes.

Early exoplanet radial-velocity (RV) detections were often confirmed by identifying significant peaks in periodograms--where false-alarm probabilities are calculated to establish the significance of \reddtext{a} planetary candidate--and performing non-linear least-squares fitting for Keplerian orbits. Such was the case for 51\,Pegasi\,b \citep{mayor95}--the first exoplanet around a Sun-like star--and for the many Jovian planets that quickly followed this discovery \citep{early_giants1, early_giants2}.

However, as RV data grew and multi-planet systems became common, more sophisticated statistical tools were needed to extract the more difficult exoplanet signals, which could be subtle, degenerate, and embedded in considerable noise. This led to the introduction of MCMC methods \citep{ford05}, revolutionising exoplanet RV fitting with robust estimation of orbital parameters with realistic uncertainties, even for non-linear parameters (like eccentricity and longitude of periastron). Far from perfect, MCMC still has its caveats. Poorly tuned chains could wander slowly or get trapped in local maxima. 

To thoroughly explore multi-modal parameter spaces (common in multi-planet systems for example) \citet{gregory_hd73526} pioneered the use of the parallel tempering (PT) MCMC method \citep{swendsen86_apt, parallel_tempering_earldeem} in relation to RV-signal detection and characterisation. This approach runs several chains in parallel to sample different powers (a temperature ladder) of the posterior distribution, each tempered to a different level. Hotter chains traverse the parameter space with more freedom, while colder chains sample the fine details. Inter-chain communication allows the sampler to explore distant high-probability nodes with ease. Furthermore, PT enables the algorithm to estimate the marginalised posterior (or evidence), a crucial quantity for model comparison. \citet{gregory_hd73526} shows that this method is capable of efficiently exploring all regions of the phase space, lessening the burden of multi-modality. However, this early implementation was not capable of estimating the evidence reliably without a huge performance loss.

The affine-invariant ensemble samplers \citep{goodman10} also found their way into the exoplanet domain \citep{hou12}. This MCMC variant, instead of a single chain, utilises an ensemble of `walkers', \reddtext{drawing multiple samples per step, and making the sampler insensitive to parameter covariances. Thus providing an enormous computational performance boost.}

As an extra Keplerian in your model can always fit noise, efforts turned to providing an accurate model comparison framework, leading to the usage of nested sampling (NS) algorithms \citep{skilling04}, to calculate the Bayesian evidence for competing models (with differing number of planets or noise models), allowing one to select the most likely model \citep{feroz08_multimodal_ns, feroz11}. Building on nested sampling, dynamic nested sampling \citep[DNS, ][]{sampler_dynamic_nested, sampler_dynesty} introduces an adaptive allocation of `sampling effort' to higher probability areas of the phase space, further increasing the efficiency of reaching the target evidence precision, finding use in RV fitting \citep{dynesty_applied_exo}.

MCMC excels in parameter posterior estimation, yielding reliable uncertainties, even for complex or correlated parameters. However, while evidence estimation is certainly possible under this algorithm, it is as a by-product of the posterior exploration. Therefore, increasing the accuracy of the evidence estimation does not necessarily translate to increased posterior accuracy, while carrying a substantial computational burden.
On the other hand, NS is primarily designed for efficient evidence estimation by systematically shrinking the likelihood volume. As a by-product, it provides posterior samples, which can be used to model the parameter posteriors. Consequently, this method inherently does not provide as many samples from the parameter distribution as a well-tuned MCMC, conveying less precise parameter estimates.

This work presents \reddemc\footnote{\url{https://reddemcee.readthedocs.io/}}, an adaptive parallel tempering MCMC \texttt{Python} algorithm for any scientific sampling related endeavour that handles complex, high-dimensional, multi-modal posteriors, with several competing models. Leveraging five different automated tuning strategies for the temperature ladder--two classic ones, uniform swap acceptance, and posterior area overlap; and three new implementations based on average energy differences, the system's specific heat, and the swap mean distance--while providing original adaptations for evidence estimation, like the stepping stones algorithm \citep[SS, ][]{steppingstones2011} with a per-stone geometric-bridge, thermodynamic integration \citep[TI, ][]{Gelman1998SimulatingNC} enhanced by curvature-aware interpolation, as well as a novel hybrid approach.

\section{Bayesian framework}  \label{sec:bayesian_framework}

A full description of Bayesian inference and MCMC methods is beyond the scope of this manuscript, and we therefore refer the reader to \citet{bayesian-data-analysis-gelman}. Nonetheless, a summary of essential concepts is provided.

Bayesian inference generally consists of depicting the posterior probability distribution over the parameters $\pmb{\theta} \in \pmb{\Theta}$ of the hypothesised model $M$, depicting some known measured data $D$:

\begin{equation} \label{eq:bayes_theorem}
    p(\pmb{\theta} \mid D, M) = \frac{p(\pmb{\theta} \mid M) \cdot p(D \mid \pmb{\theta}, M)}{p(D \mid M)} \,,
\end{equation}

where $p(\pmb{\theta} \mid M)$ corresponds to the prior distribution; $p(D \mid \pmb{\theta}, M)$ to the likelihood function; and $p(D \mid M)$ to the marginalised likelihood, frequently denoted as the Bayesian evidence $\mathcal{Z}$, or simply `evidence' (as it will be in this manuscript). Throughout this work, proper priors are assumed ($\int_{\Theta} p(\pmb{\theta} \mid M)=1$), so the evidence is well-defined and finite.

The evidence quantifies the overall support for model M and is crucial for Bayesian model comparison. However, the evidence integral generally has no closed-form solution and must be estimated numerically \citep{oaks_2018}. MCMC methods, while excelling at drawing samples from the posterior, do not directly provide $\mathcal{Z}$.

Within MCMC methods, several evidence-estimation techniques have been developed \citep{maturana18}. For parallel tempering MCMC, two common approaches are thermodynamic integration \citep{Gelman1998SimulatingNC} and stepping stones \citep{steppingstones2011}. How parallel tempering works and how TI and SS leverage the tempered ensemble to estimate the evidence are briefly outlined below.

\subsection{Parallel tempering MCMC} \label{sec:parallel_tempering}

PT MCMC runs an ensemble of $B$ parallel chains, each sampling a different power of the posterior distribution (at a different temperature $T_i$; \citealp{swendsen86_apt, parallel_tempering_earldeem, miasojedow_12}). The inverse of the temperature is $\beta = \frac{1}{T}$, with $1 = \beta_1 > \beta_2>...>\beta_{B-1} >\beta_B \geq 0$, and the full chain $\pmb{\bar{X_t}} = (X_t^{(\beta_1)}, ..., X_t^{(\beta_B)})$. By convention, $\beta_1=1$ for the cold chain (sampling the original posterior), and $\beta_{B} \approx 0$ for the hottest chain (sampling a highly flattened landscape, approaching the prior as $\beta_B\xrightarrow{}0$). Each chain $X_t^{(\beta_i)}$ samples a tempered posterior where the likelihood has been raised to the power $\beta_i$

\begin{equation} \label{eq:bayes_theorem_tempered}
    \begin{aligned}
    \mathcal{P}_{\beta_i}(\pmb{\theta}) &= \frac{\Pi \cdot \mathcal{L}^{\beta_i}}{\mathcal{Z}_{\beta_i}}\,,
    \end{aligned}
\end{equation}

where $\mathcal{P}_{\beta_i}(\pmb{\theta})$ is the posterior distribution , $\Pi$ the prior, $\mathcal{L}^{\beta_i}$ the tempered likelihood, and $\mathcal{Z}_{\beta_i}$ the evidence.

In $\mathcal{P}_{\beta}(\pmb{\theta})$, hotter temperatures flatten and broaden peaks, reducing the risk of getting trapped in local maxima, and effectively making the posterior easier to sample. As such, hot chains are able to rapidly sample a large portion of the parameter space, whilst cold chains provide precise local sampling.
By allowing chains at different temperatures to swap states, PT achieves a fast, thorough sampling of the posterior. Swaps in PT are implemented with a Metropolis-like acceptance rule. The pairwise swap \citep{goggans_2004}, which is when two adjacent chains $(X_t^{(\beta_i)}, X_t^{(\beta_{i+1})})$ propose to switch states with probability

\begin{equation}  \label{eq:swap_acceptance_rate}
    A_{i, i+1} = \mathrm{min}\left(0, -\Delta \mathcal{\ln{L}}^{\beta_i} \cdot \Delta \beta_i \right)\,,
\end{equation}

where $\Delta \mathcal{\ln{L}}_i$ is the log-likelihood difference between the two chain states. This criterion (derived by simplifying the detailed balance condition for swapping two Boltzmann-distributed ensembles) shows that swaps are more probable when a high-likelihood state from a hotter chain is proposed for a colder chain. Without adjacent posteriors overlapping enough to allow regular swaps, efficiency plummets. By tuning the temperature ladder (the set of $\beta_i$ values), PT aims to maintain a reasonably high swap-acceptance rate across all adjacent pairs, ensuring a thorough exploration of the posterior landscape and providing a natural framework for estimating the evidence $\mathcal{Z}$ via the ladder of tempered distributions.

Although our APT implementation does not provide independent draws, we next set up the classic methods for evidence estimation (which assume IID-based errors). In \refsec{sec:revisiting_evidence}, we address how to handle this problem.

\subsection{Thermodynamic integration} \label{sec:ti}

Thermodynamic integration is an indirect method to compute the evidence $\mathcal{Z}$ by leveraging the continuous sequence of intermediate posteriors between prior and posterior \citep{Gelman1998SimulatingNC, goggans_2004, lartillot06}. 
The  basic formula arises from the identity (in statistical mechanics) that relates the derivative of the log-partition function to the average energy. In Bayesian terms:

\begin{equation} \label{eq:evidence_ti0}
    \ln{\mathcal{Z}} = \ln{\mathcal{Z}_1} - \ln{\mathcal{Z}_0}= \int_0^1 \mathbb{E}_{\beta} \left[ \ln{\mathcal{L}}\right]d\beta \,,
\end{equation}

where $\mathbb{E}_{\beta} \left[ ...\right]$ is the expectation with respect to $\mathcal{P}_{\beta}(\pmb{\theta})$, the tempered posterior at inverse temperature $\beta$. Since $\mathcal{Z}_0=1$ for proper priors (so $\ln{\mathcal{Z}_0}=0$), the integral directly yields $\ln{\mathcal{Z}}$. With $B$ the number of temperatures, the integral is calculated via the trapezoidal rule:

\begin{equation} \label{eq:evidence_ti}
\ln{\widehat{\mathcal{Z}}}_{\mathrm{TI}} \approx \sum_{i=1}^{B-1} \frac{\mathbb{E}_{\beta_{i + 1}} \left[ \ln{\mathcal{L}}\right] + \mathbb{E}_{\beta_{i}} \left[ \ln{\mathcal{L}}\right]}{2} \cdot\Delta\beta_i\,.
\end{equation}

\subsection{Stepping stones}

The stepping-stones algorithm \citep{steppingstones2011} expresses the evidence ratio $\mathcal{Z}$=$\mathcal{Z}_1/\mathcal{Z}_0$ as a telescopic product across the temperature ladder:

\begin{equation} \label{eq:ss_ratios}
\widehat{\mathcal{Z}}_{\mathrm{SS}} = 
\prod_{i=1}^{B-1}\frac{\mathcal{Z}_{\beta_i}}{\mathcal{Z}_{\beta_{i+1}}} = 
\prod_{i=1}^{B-1} r_i\,,
\qquad
r_i \equiv \frac{\mathcal{Z}_{\beta_i}}{\mathcal{Z}_{\beta_{i+1}}} = \mathbb{E}_{\beta_{i+1}}\left[ \mathcal{L}^{\beta_i-\beta_{i+1}}\right]\,,
\end{equation}

with $r_i$ the stepping-stones ratios and $\mathbb{E}_{\beta} \left[ ...\right]$ the expectation with respect to $\mathcal{P}_{\beta}(\pmb{\theta})$. It is often convenient to work in log-space for numerical stability, yielding

\begin{equation} 
    \ln \widehat{\mathcal{Z}}_{\mathrm{SS}} = \sum_{i=1}^{B-1} \ln\left[\frac{1}{N_i}\sum_{n=1}^{N_i} \mathcal{L}^{\beta_i-\beta_{i+1}}\right]\,,
\end{equation}

where $N_i$ is the number of posterior samples per chain used to approximate each expectation. SS has the advantage of typically delivering improved accuracy over TI when the number of temperatures is limited, since it more directly `bridges' distributions to evaluate each incremental evidence ratio.

\subsection{Evidence error estimation} \label{sec:evidence_error_est}

For TI, \citet{lartillot06} identify two sources of error: one from the sampling itself $\widehat{\sigma}_S$, calculated as the MC standard error affecting the log-likelihood averages, and another from the discretisation of the integral $\widehat{\sigma}_D$. This quantity can be estimated by considering the worst-case contribution of the trapezoidal rule (essentially misrepresenting the whole triangular part):

\begin{equation} \label{eq:zerr_disc}
    \widehat{\sigma}_D \simeq \frac{|\mathbb{E}_{\beta_{i}}[\ln\mathcal{L}] - \mathbb{E}_{\beta_{i+1}}[\ln\mathcal{L}]|}{2}\cdot \Delta\beta_i \,.
\end{equation}

For the SS method, the product of unbiased estimators $\hat{\mathcal{Z}}_{\mathrm{SS}}$ is also unbiased, nevertheless, changing to log scale introduces a bias. \citet{steppingstones2011} estimate the error as
\begin{equation} \label{eq:z_err_ss}
    \widehat{\sigma}^2_{\mathcal{Z}_\mathrm{SS}} \approx \frac{1}{N^2} \sum_{i=1}^{B-1} \sum_{n=1}^N \left( \frac{\mathcal{L}^{\beta_i-\beta_{i+1}}}{\hat{r}_i} -1 \right)^2\,.
\end{equation}

\section{Temperature ladder} \label{sec:temp_ladder}

Extending the parallelisms with statistical thermodynamics, the specific heat $C_{\nu}$ is defined with the energy $U=-\ln \mathcal{L}$:
\begin{equation}
    C_{\nu}(\beta) = \frac{d\mathbb{E}_{\beta}[U]}{dT} = \beta^2 (\mathbb{E}_{\beta}[U^2] - \mathbb{E}_{\beta}[U]^2)
    = \beta^2\mathrm{Var}_{\beta}\left[U \right]\,.
\end{equation}
This quantity governs the energy fluctuations at different temperatures, and is closely related to both the temperature ladder and the swap-acceptance rate.
The mean swap-acceptance rate (SAR)--derived from \refeq{eq:swap_acceptance_rate}--for a small temperature gap can be expressed as

\begin{equation}  \label{eq:sar}
\begin{aligned}
    \bar{A}_{i, i+1} &= \mathbb{E}_{\beta_i, \beta_i+1}\left[\mathrm{exp}(-\Delta U_{i} \cdot \Delta\beta_i ) \right] \\ &\propto \mathbb{E}_{\beta_i}\left[\Delta U_i \cdot \Delta \beta_i\right] = \frac{\sqrt{C_{\nu}(\beta_i)}}{\beta_i} \Delta \beta_i\,.
\end{aligned}
\end{equation}

In the PT scheme, the temperature ladder must be chosen to maximise the efficiency of the cold chain. To ensure this chain has good mixing, from the SAR (see \refeq{eq:sar}), it would suffice to minimise both $\Delta \beta_i$ and $\Delta U_i$ for every single chain. This way, samples from the hot chain can easily traverse to the cold chain.
As a secondary aim, the temperature ladder serves to estimate the evidence as well, so allocating the temperatures aimed at favouring either the TI or SS method also has its benefits.  Achieving a Uniform SAR across all chains is generally considered a good strategy for healthy mixing \citep{sugita99, predescu2004_beta_law, kone2005, roberts2007, ptemcee}. 

\subsection{Geometric spacing} \label{sec:temp_ladder:fixed}
\citet{kofke2002} and later \citet{predescu2004_beta_law}, by analysing the constant specific heat scenario, conclude that adjacent temperatures geometrically spaced should approximate to a uniform SAR. With a constant ratio $\bar{R}$, each $\beta$ would be defined by
\begin{equation}
    \bar{R} = \frac{\beta_i}{\beta_{i+1}}\,.
\end{equation}

This is the starting point for ladder design, and it illustrates why adaptive methods are needed: geometric spacing assumes constant $C_{\nu}$, which does not hold true in complex posteriors.

\subsection{Adaptive methods} \label{sec:temp_ladder:adaptive}

Adaptive parallel tempering (APT) MCMC dynamically updates the \reddtext{temperature ladder during} the run to improve mixing efficiency \citep{gilks98_apt, roberts2007, miasojedow_12, ptemcee}. Such an algorithm inherently fails to be ergodic and may not preserve the stationarity of the cold chain.
In practice, \reddemc~uses a decaying adaptive rate (see \refeq{eq:kappa}), allowing the ladder to stabilise over time. Then, the adaptation can be discontinued, after which detailed balance is preserved and standard ergodicity theorems apply. This procedure \reddtext{was} adopted for all benchmarks that follow. Next, five different adaptation strategies are shown, of which the first two--uniform swap-acceptance rate and Gaussian area overlap--are commonly seen in the literature, while the latter three--swap mean distance, small Gaussian gap, and equalised thermodynamic length--we present our implementations, as \reddtext{effective} alternatives, especially when $C_\nu$ is non-uniform.

\subsubsection{Uniform swap-acceptance rate} \label{sec:temp_ladder:adaptive:uni_a}

\citet{ptemcee} propose that the dynamic adjustments of the temperature ladder are guided by the SAR. They define the ladder in terms of logarithmic temperature intervals $S_i$ between adjacent chains. Therefore, the adaptation rate corresponds to $\frac{dS_i}{dt}$. The adaptation includes a diminishing factor $\kappa(t)$, where $\lim_{t\to\infty} \kappa(t) = 0$, here expressed as a hyperbolic decay:

\begin{align} 
    S_i &\equiv \ln{(T_{i} - T_{i+1})}\,,
    \qquad \frac{dS_i}{dt} = \kappa(t)\left[ A_i(t) - A_{i+1}(t)\right]\,, \label{eq:dsdt} \\
    \kappa(t) &= \frac{1}{\nu_0}\frac{\tau_0}{(t + \tau_0)}\,, \label{eq:kappa}
\end{align}

with $A_i(t)$ the measured SAR, $\tau_0$ the decay half-life, and $\nu_0$ the evolution time-scale. This leaves both $\beta_1$ and $\beta_B$ static, while the intermediate temperatures settle to achieve a uniform SAR.

\subsubsection{Gaussian area overlap} \label{sec:temp_ladder:adaptive:gauss_area_overlap}
\citet{rathore2005}, by studying the correlation between replicas of the SAR and the area of overlap of likelihoods in Gaussian distributions, found a relation to control the acceptance rate by spacing temperatures:

\begin{equation}
    A_{\mathrm{overlap}} = \mathrm{erfc}\left[ \frac{\Delta \ln{\mathcal{L}}}{2\sqrt{2}\sigma_m}\right] \quad \Rightarrow \quad  \frac{\Delta \ln{\mathcal{L}}}{\sigma_m}\Bigg|_{\beta_i} = \left[ \frac{\Delta \ln{\mathcal{L}}}{\sigma_m}\right]_{\mathrm{target}}\,,
\end{equation}

where $\sigma_m = (\sigma_1+\sigma_2)/2$ is the mean of the deviations of the adjacent temperature log-likelihoods. While also aiming for Uniform SAR, this ladder evolution is informed by a different mechanism, which may compensate the effects of having a non-uniform $C_{\nu}$.

\subsubsection{Uniform swap mean distance} \label{sec:temp_ladder:adaptive:uni_smd}
Considering that a non-uniform $C_{\nu}$ is more realistic for complex systems (as in exoplanet discovery), \citet{katzgraber2006} present a feedback-optimised method, which maximises the round trips of each replica between the extremal temperatures. This allows information about the likelihood landscape to travel faster from the hot chain to the cold chain.
A well-tempered chain (where the posterior shape has changed significantly compared to the target posterior) should propose far-away states which, if accepted, would contribute more to healthy mixing than a swap so close that it is equivalent to the intra-chain evolution.
Therefore, a ladder where adjacent swaps are done by maximising the swap distance would maximise the crossing of `useful' information across the replicas.
By normalising all dimensions in the system to unity, the mean distance of adjacent swaps for all walkers at a given temperature is

\begin{equation}
d|_{\beta_i} = \frac{1}{W}\sum_{w=1}^\mathcal{\mathrm{W}}\sqrt{\sum_{d=1}^{\mathrm{D}}\left(\frac{\theta_{d, i}-\theta_{d, i+1}}{c_d}\right)^2}\,,
\end{equation}

where $W$ is the number of walkers, $D$ the number of dimensions, and $c_d$ the range of the prior for dimension $d$. Hereafter this method is referred to as the swap mean distance ($\mathrm{SMD}$). Intuitively, this encourages swaps that carry proposals farther across the parameter space, which may be specifically beneficial in high-dimensional problems, where local swaps might exchange nearly similar states.

\subsubsection{Small Gaussian gap} \label{sec:temp_ladder:adaptive:sgg}
This scheme (SGG) targets a uniform SAR by using a Gaussian approximation for small temperature gaps. If adjacent $\beta$s are very close, the energy difference $\Delta U_i$ has approximately a Gaussian distribution $\sim$$\mathcal{N}(0, 2\mathrm{Var}[U_i])$. In this small-gap regime, one can show that the product $(\Delta \beta_i)^2 \mathrm{Var}[U_i]$ largely controls the swap-acceptance rate \citep{predescu2004_beta_law}, from \refeq{eq:sar}:

\begin{equation} \label{eq:sar2}
    \begin{aligned}
    \bar{A}_{i, i+1} &= \mathbb{E}_{\beta_i, \beta_i+1}\left[\mathrm{exp}(-\Delta U_{i} \cdot \Delta\beta_i ) \right]  \\
    &\approx \mathrm{exp}\left(\frac{1}{2} (\Delta \beta_i)^2(2\mathrm{Var}_{\beta_i}[U])\right) =\mathrm{exp}\left((\Delta \beta_i)^2(\mathrm{Var}_{\beta_i}[U])\right)\,.
    \end{aligned}
\end{equation}

Thus, keeping the product $(\Delta \beta_i)^2(\mathrm{Var}[U_i])$ constant \reddtext{implies} a roughly uniform SAR.

\subsubsection{Equalised thermodynamic length} \label{sec:temp_ladder:adaptive:thermodynamic}

The thermodynamic length is defined as
\begin{equation} \label{eq:thermodynamic_length}
    L(\beta) = \int_{\beta_{\mathrm{min}}} ^{\beta_{\mathrm{max}}}\sqrt{\mathrm{Var_{\beta}[U]}} d\beta = \int_{\beta_{\mathrm{min}}} ^{\beta_{\mathrm{max}}} \frac{\sqrt{C_{\nu}}(\beta)}{\beta} d\beta\,.
\end{equation}

Intuitively $L(\beta)$ corresponds to the local metric which measures how quickly (or far) the distribution changes with $\beta$. Setting neighbouring replica' distributions so they are the same `distance' apart translates to placing them uniformly in $L$-space, effectively populating the regions where $C_{\nu}$ peaks with more chains. This ensures denser sampling where the posterior changes most rapidly.
Each interval of this integral can be approximated as
\begin{equation}
    \Delta L_i \approx \frac{\Delta \beta_i}{2} (\sqrt{\mathrm{Var_{\beta_i}[U]}} +\sqrt{\mathrm{Var_{\beta_{i+1}}[U]}})\,.
\end{equation}

Then, the discrete estimate for thermodynamic distance is
\begin{equation}
    L(\beta_N) = \sum_{i=0}^{B-1}\Delta L_i \,.
\end{equation}

\citet{shenfeld09} arrives at the same conclusion, further demonstrating that under constant $C_{\nu}$, the ETL forms a geometric progression as well.

\subsubsection{Ladder adaptation in \reddemc} \label{sec:temp_ladder:implementation}

\reddemc~implements each of the methods discussed above by setting the adaptation as $\frac{dS_i}{dt}$=$\kappa(t) \cdot Q_i$. This way, choosing a definition for $Q_i$ determines the adaptive approach--be it uniform swap-acceptance rate (SAR), Gaussian area overlap (GAO), uniform swap mean distance (SMD), small Gaussian gap approximation (SGG), or equalised thermodynamic length (ETL):

\begin{equation}
  Q_i =
  \begin{cases}
    \left[ A_i(t) - A_{i+1}(t)\right] & \text{for SAR (see \refsec{sec:temp_ladder:adaptive:uni_a}}) \\
    \frac{\Delta U_i}{\sigma_m} &    \text{for GAO (see \refsec{sec:temp_ladder:adaptive:gauss_area_overlap}})
    \\
    d|_{\beta_i} & \text{for SMD (see \refsec{sec:temp_ladder:adaptive:uni_smd}})\\
    \mathrm{exp}\left((\Delta \beta_i)^2(\mathrm{Var}[U_i])\right) & \text{for SGG (see \refsec{sec:temp_ladder:adaptive:sgg}})\\
    \frac{\Delta L_i}{L(\beta_N)} & \text{for ETL (see \refsec{sec:temp_ladder:adaptive:thermodynamic}})
  \end{cases}\,.
\end{equation}

\subsection{Revisiting evidence} \label{sec:revisiting_evidence}
The APT implementation does not provide independent draws; each walker's proposal is based on its current position, meaning successive samples retain memory until the integrated autocorrelation time has passed. The stretch move also couples walkers, influencing each other's proposals \citep{goodman10}. Furthermore, the swap move creates a cross-temperature coupling as well. To account for all these effects in correlated draws, we replace IID-based errors with autocorrelation-adjusted MC standard errors \citep{flegal08_bm_sde}, using overlapping batch means \citep[OBM, ][]{wang_2017_obm}. This technique, which is consistent for MCMC under standard ergodicity conditions, also propagates cross-temperature covariance across the ladder. We use this as $\widehat{\sigma}_S$ for both methods.

Both TI and SS rely on the shape of the $\mathbb{E}_{\beta}[U](\beta)$ curve, for which we consider three factors:
1) the ladder size is relatively small, providing few samples for either TI or SS; 2) temperatures are not uniformly spaced; 3) $\mathbb{E}_{\beta}[U](\beta)$ is a monotonically increasing function.

With these points in mind, two different solutions for a better $\widehat{\sigma}_D$ come to mind. One approach is to use the Richardson extrapolation, which entails computing the evidence estimate on two different grids (one coarse and one fine). The difference between these estimates is then used to extrapolate the discretisation error, leveraging the known convergence order of the trapezoidal rule. Another approach is to estimate the local curvature of the $\mathbb{E}_{\beta}[U](\beta)$ curve (since the leading trapezoidal-rule error term is proportional to the second derivative) to derive a tailored error estimate for each interval. In the following paragraphs two original implementations based on these ideas are presented, leveraging the properties of $\mathbb{E}_{\beta}[U](\beta)$ for a more precise $\ln \mathcal{Z}$ estimation.

\paragraph{Piecewise interpolated thermodynamic integration}
Building on the TI method, an interpolation on the curve $\mathbb{E}_{\beta}[U](\beta)$ is applied with the local Piecewise Cubic Hermite Interpolating Polynomial (PCHIP), which preserves monotonicity (physically expected) and prevents overshooting for non-smooth data (few temperatures; \citealp{pchip}).

For $\widehat{\sigma}_D$, we form a coarse ladder by dropping every other temperature, build the corresponding PCHIP, and compute another coarser evidence estimate. We use its difference with respect to the finer one as the discretisation error.

The total error is then $\widehat{\sigma}_{\mathcal{Z}} = \sqrt{\widehat{\sigma}_D^2+\widehat{\sigma}_S^2}$. It is worth noting that this approach addresses the discretisation error more explicitly, thus reducing bias at the cost of possibly smaller (but more realistic) error bars, as shown in Section~\ref{sec:benchmarks}. More details on the method can be found in \ref{app:ti}.

\paragraph{Geometric-bridge stepping stones}

The SS evidence estimator $\hat{\mathcal{Z}}_{\mathrm{SS}}$ (see \refeq{eq:z_err_ss}) estimates the ratio $r_i$ with an expectation under a single distribution $\mathbb{E}_{\beta_i+1}[\cdot]$ (see \refeq{eq:ss_ratios}). We replace the ratios with geometric-bridge sampling \citep{meng96_bridge_sampling, gronau17_bridge_sampling}:

\begin{equation}
    r_i=\frac{\mathbb{E}_{\beta_{i}}\left[ \mathcal{L}^{\Delta\beta_i/2}\right]}{\mathbb{E}_{\beta_{i+1}}\left[ \mathcal{L}^{-\Delta\beta_i/2}\right]}\,.
\end{equation}

By using samples from both ends, we expect to diminish the estimator variance when the temperature gap is large, as the per-stone ratio would degrade less rapidly. Using the same APT samples has no additional computational cost, and by halving the exponent we reduce the dynamic range, improving numerical stability. A more detailed explanation can be read in \refapp{app:ss}.

\paragraph{Hybrid method}

Consider a temperature mid-point $\beta_*$, so thermodynamic integration is applied in $\beta \in [0, \beta_*]$, and stepping stones in $\beta \in [\beta_*, 1]$. Provided that continuity is ensured around $\beta_*$, separating where each method is stronger leads to a fair estimation of $\ln\mathcal{Z}$. From \refeq{eq:evidence_ti} the overall log-evidence can be decomposed as
\begin{equation}
\ln\mathcal{Z} = \int_0^1 \mathbb{E}_{\beta} \left[{U}\right] d\beta =
\int_0^{\beta_*} \mathbb{E}_{\beta} \left[{U}\right] d\beta + \int_{\beta_*}^1 \mathbb{E}_{\beta} \left[{U}\right] d\beta\,.
\end{equation}

The transition between the `pure prior' region ($\beta$=$0$) to a moderately informed distribution (small $\beta$) changes statistics more dramatically than other areas. Intuitively, any sharp change in posterior shape can be seen as a phase transition, which is also reflected as a peak in $C_{\nu}(\beta)$. Since most temperature ladders are proportional in some form to $C_{\nu}$, chains are denser in this region, and therefore, a straightforward method like TI works well.
Near $\beta$=$1$, SS is more efficient at handling ratios since the posterior is strongly peaked and $\Delta\beta$ tends to increase. By decomposing the TI trapezoidal area contribution of each interval, the mid-point is selected to be where the rectangular area is larger than twice the triangular area, ensuring we are in a region where $\Delta U_i$ dominates over $\Delta \beta_i$, and where the energy slope is pronounced:

\begin{equation}
    \Delta\beta_i \cdot U_i \geq 2(\Delta\beta_i \cdot \frac{\Delta U_i}{2}) \Rightarrow 2U_i \geq U_{i+1}\,.
\end{equation}

\section{Benchmarks}  \label{sec:benchmarks}

This section seeks to validate the proposed ladder adaptation algorithms as well as the evidence estimators. Taken together, these tests probe the different problems that matter most when contrasting an APT sampler with DNS. Gaussian shells, Gaussian egg-box, and the hybrid Rosenbrock function will capture the three classic challenges when sampling from a multimodal distribution: mode-finding, mode-hopping, and in-mode mixing, all of which are intensified as dimensionality increases.
Hereafter, two distinct qualities are compared: first, the sampler performance (comprising both the intra-chain mixing and the computational cost, measured in effective samples per second or `kenits'); and second, the evidence estimation (comprising the estimated evidence value and its uncertainty). This will be measured by both the difference found with the analytical evidence \zdiff$\equiv$$\exp \left( \ln{\mathcal{Z}}-\ln\widehat{\mathcal{Z}}\right)-1$, validating the accuracy of the estimator, and the log-likelihood \zlogl$\equiv$$\ln\mathcal{L}_{\widehat{\mathcal{Z}}}$, validating the credibility of the uncertainties, for $\ln\widehat{\mathcal{Z}}$$\sim$$\mathcal{N}(\ln\widehat{\mathcal{Z}}, \widehat{\sigma}_{\ln{\widehat{\mathcal{Z}}}})$, where $\ln\widehat{\mathcal{Z}}$ is the estimate of the evidence, and $\widehat{\sigma}_{\ln{\widehat{\mathcal{Z}}}}$ the estimate of the evidence uncertainty.

Each benchmark is run 11 times with different known random seeds, and the reported values correspond to the mean and standard deviation of all runs.

\subsection{n-d Gaussian shells}\label{sec:benchmarks_gaussian_shells}

\begin{figure}
    \includegraphics[width=\columnwidth]{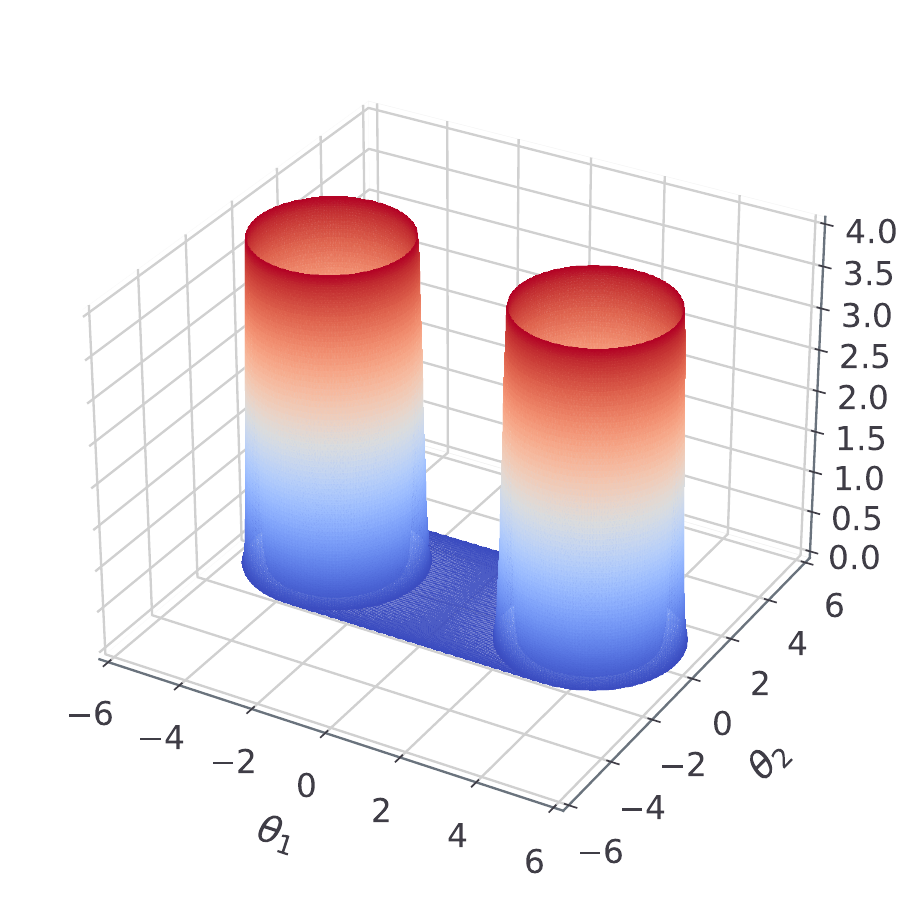}
    \caption{\label{fig:2dg:model}2-d Gaussian shells likelihood. Radius $r$=$2$ and width $w$=$0.1$. Parameter boundaries are imposed as $\pm6$. Likelihood values are coloured from blue (low) to red (high) to facilitate the visualisation of the figure.}
\end{figure}

Sampling from an n-dimensional Gaussian shell is an analytically tractable problem widely used in the literature\citep{feroz08_multimodal_ns, sampler_dynesty}. The likelihood contours are curved, thin, and nearly singular (see \reffig{fig:2dg:model}). Random-walk steps are inefficient, so this multimodal function stresses proposal geometry and temperature allocation. \reddtext{For $n$ dimensions, radius $r$, width $w$, and peaks' centres $\vec{c_i}$, the likelihood is}

\begin{equation}\label{eq:gaussian_shells}
    p(\vec{\theta}) = \sum_{i=1}^n \frac{1}{\sqrt{2\pi w^2}} \exp{\left( -\frac{(|\vec{\theta} - \vec{c_i}| - r)^2}{2w^2} \right)}\,. 
\end{equation}

This benchmark is used to assess the behaviour of the different samplers in efficiency and evidence estimation. \reddemc~was set up with $[16, 320, 640, 1]$ temperatures, walkers, sweeps, and steps, respectively, for a total of 3\,276\,800 samples. The adaptation hyper-parameters $\tau_0, \nu_0$ are set as one-tenth of the sweeps and one-hundredth of the walkers, respectively. We adapt the temperature ladder during the first half of the sweeps, and then freeze it. This entire adaptive phase is treated as burn-in: those samples are discarded before any statistical inference, but the time spent is included in the reported run-time (i.e., it affects the kenits metric). For comparison, the DNS runs were performed using \dynesty~\citep{sampler_dynesty} with its automated settings, testing three sampling modes: uniform, slice, and random-slice.

For \dynesty, we calculated kenits using the reported number of likelihood calls, effectively measuring the sampling efficiency (fraction of independent samples) times all likelihood evaluations divided by the total run-time. For \reddemc, efficiency was defined as the inverse of the average autocorrelation time of the cold-chain, and kenits were computed as this efficiency multiplied by the number of likelihood evaluations (discarding burn-in samples, but not their run-time) over the full run-time. This method ensures a fair kenits comparison.

\paragraph{Sampler performance} All adaptive ladder methods are statistically similar (see \hyperref[tab:2dgs_nits]{Table \ref{tab:2dgs_nits}}); variance between ladder algorithms ($0.04^2$) is smaller than the run-to-run variance of each method ($\approx$$0.12^2$), delivering $\approx$$3.6$-$3.8$ kenits. The best DNS method, dyn-rs, achieves $0.94$$\pm$$0.05$ kenits, which is $\sim$26\% of the APT average.
This method also finishes the fastest, at $\sim$13.4s. It is worth mentioning that once converged, the efficiency does not radically change, so even by demanding a higher total iteration count as a stop condition, the kenits (and efficiency) would not dramatically increase. The dynamic algorithm stops at target precision, which, in the dyn-u case, required a much longer run-time with many more samples, whereas the slice methods finished faster with far fewer independent samples. In other words, there is a trade-off between thorough posterior sampling and evidence-estimation convergence speed. The same reasoning can be extended to MCMC, preserving much richer posterior information.

\begin{table}
    \caption{\label{tab:2dgs_nits}2-d Gaussian shells performance benchmark.}
    \centering
        \begin{tabular}{llll}
            \toprule
            Method      & Time (s) & Eff (\%)   &   Kenits     \\
            \midrule
            \midrule
            SAR         &37.60±0.28&   8.49±0.27&   3.70±0.12  \\
            SMD         &37.77±0.19&   8.59±0.27&   3.73±0.13  \\
            SGG         &37.86±0.15&   8.49±0.30&   3.67±0.13  \\
            GAO         &37.95±0.25&   8.34±0.25&   3.60±0.11  \\
            ETL          &37.94±0.22&   8.49±0.24&   3.66±0.10  \\
            \midrule
            dyn-u       &68.66±16.75&  4.94±0.82&   0.19±0.04  \\
            dyn-s       &16.83±0.89&   3.77±0.22&   0.75±0.04  \\
            dyn-rs      &13.41±0.64&   4.49±0.33&   0.94±0.05  \\

            \bottomrule
        \end{tabular}
        
    \tablefoot{\reddemc's adaptive algorithms compared to \dynesty's uniform (dyn-u), slice (dyn-s), and random-slice (dyn-rs) sampling methods. From left to right: Time--total run time in seconds, Eff--sampling efficiency or percentage of independent samples, and kenits--effective samples per second in thousands.}
\end{table}

\paragraph{Evidence estimation} The proposed evidence estimation methods in \refsec{sec:evidence_error_est} are tested and compared against the analytical value for the 2-d case, $\ln{\mathcal{Z}}$=$-1.746$ (see \hyperref[tab:2dgs_zest]{Table \ref{tab:2dgs_zest}}). 
The default TI method has \zdiff$|_{\mathrm{TI}}$=$8.394\%$, whereas the improved TI+ version achieves \zdiff$|_{\mathrm{TI+}}$=$0.324\%$ (an order of magnitude closer to the true value). The likelihood improves as well, with a ratio of $\exp(2.80-0.92)=6.58$. By contrast, the default SS has the highest \zlogl$|_{\mathrm{SS}}$=$4.326$ alongside a low \zdiff$|_{\mathrm{SS}}$=$0.117\%$. The SS+ counterpart shows slightly more conservative uncertainties (0.006 instead of 0.005) with \zlogl$|_{\mathrm{SS+}}$=$4.263$, and slightly better accuracy \zdiff$|_{\mathrm{SS+}}$=$0.099\%$.
The H+ method achieves the highest accuracy of all \zdiff$|_{\mathrm{H+}}$=$0.016\%$, along a high likelihood as well \zlogl$|_{\mathrm{H+}}$=$3.951$. Out of the DNS methods, dyn-u is the most reliable with \zdiff$|_{\mathrm{dyn-u}}$=$1.751\%$, and a log-likelihood of \zlogl$|_{\mathrm{dyn-u}}$=$2.206$.

\begin{table}
    \centering
    \caption{\label{tab:2dgs_zest} 2-d Gaussian shells SAR evidence estimation comparison.}
        \begin{tabular}{lllll}
            \toprule
            Method      & \lzest      & \lzerrest  & \zdiff (\%) & \zlogl\\
            \midrule
            \midrule
            TI          &-1.833±0.008 &0.066±0.005 &8.394   &0.917   \\
            SS          &-1.744±0.008 &0.005±0.001 &\textbf{0.117}   &\textbf{4.326}   \\
            H           &-1.788±0.008 &0.134±0.005 &4.124   &1.044   \\
            \midrule
            TI+         &-1.749±0.008 &0.024±0.005 &0.324   &2.801   \\
            SS+         &-1.745±0.008 &0.006±0.001 &0.099   &\textbf{4.263}   \\
            H+          &-1.745±0.008 &0.008±0.002 &\textbf{0.016}   &3.951   \\
            \midrule
            dyn-u       &-1.763±0.054 &0.040±0.001& \textbf{1.751}&    \textbf{2.206} \\
            dyn-s       &-1.774±0.055 &0.040±0.001& 2.876&    2.049 \\
            dyn-rs      &-1.767±0.043 &0.040±0.001& 2.159&    2.157 \\

            \bottomrule
        \end{tabular}
        
    \tablefoot{\reddemc's adaptive algorithms compared to \dynesty's uniform (dyn-u), slice (dyn-s), and random-slice (dyn-rs) sampling methods. From left to right, the log-evidence estimator, the estimator uncertainty, the difference to the true value $\ln{\mathcal{Z}}$=$-1.746$ in percentage \zdiff, and the log-likelihood of the estimator \zlogl.}

\end{table}

\subsection{Gaussian egg-box}\label{sec:benchmarks_eggshells}
\begin{figure}
    \includegraphics[width=\columnwidth]{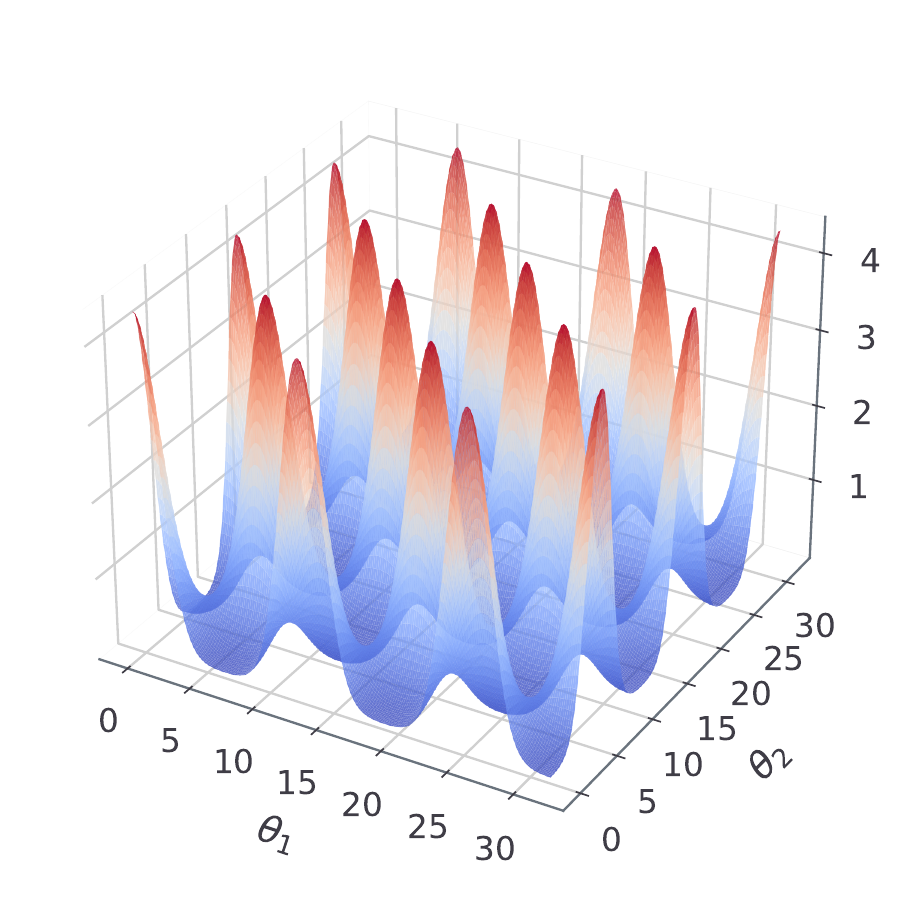}
    \caption{\label{fig:eggbox}Gaussian egg-box likelihood, with 16 different modes. Likelihood values are coloured from blue (low) to red (high) to facilitate visualisation.}
\end{figure}

The Gaussian egg-box, dubbed after its likelihood shape (see \reffig{fig:eggbox}), presents a periodic landscape with several equally high peaks, separated by deep troughs. It is an ideal stress test for mode completeness. Missing a single peak would render a catastrophically wrong evidence. 
The likelihood function is 
\begin{equation} \label{eq:eggbox}
        p(\vec{\theta}) = (2 + \prod_{n=1}^{dim} \cos{(\frac{\theta}{4})})^{\bar{\beta_{e}}}\,,
\end{equation}
where $\bar{\beta_{e}}=5$ is used, defining the narrowness of the peaks. The prior volume is limited to [0, 10$\pi$], and a fine grid integration gives $\ln{\mathcal{Z}}$=$235.856$, which will be used as the true value. We use the same setup as in \refsec{sec:benchmarks_gaussian_shells}.

\paragraph{Sampler performance} For the ladder adaptation methods, the SAR and SMD methods rank best at $5.79\pm0.29$ and $5.66\pm0.23$ kenits with overlapping uncertainties, followed closely by the other methods (see \hyperref[tab:eggbox_enits]{Table \ref{tab:eggbox_enits}}). Out of the DNS methods, dyn-rs has the lowest wall-time as well as the highest kenits count $0.82 \pm 0.07$, a $14.2\%$ of the SAR kenits.

\begin{table}
    \centering
    \caption{\label{tab:eggbox_enits} Egg-box performance benchmark.}
        \begin{tabular}{llll}
            \toprule
            Method          & Time (s)& Eff (\%)   & Kenits     \\
            \midrule
            \midrule
            SAR         & 22.90±0.11&   8.10±0.43&   5.79±0.29   \\
            SMD         & 23.02±0.24&   7.95±0.32&   5.66±0.23   \\
            SGG             & 23.35±0.14&   7.69±0.28&   5.40±0.19   \\
            GAO             & 23.36±0.24&   7.84±0.30&   5.50±0.20   \\
            ETL              & 23.52±0.26&   7.90±0.38&   5.51±0.28   \\
            \midrule
            dyn-u       & 30.41±1.51&  10.71±0.88&   0.45±0.02 \\
            dyn-s       & 19.83±0.79&  3.06±0.22 &   0.69±0.03 \\
            dyn-rs      & 16.87±1.50&  3.21±0.39 &   0.82±0.07 \\
            \bottomrule
        \end{tabular}
        
    \tablefoot{\reddemc's adaptive algorithms compared to \dynesty's sampling methods. From left to right: Time--total run time in seconds, Eff--sampling efficiency or percentage of independent samples, and kenits--effective samples per second in thousands.}
\end{table}

\paragraph{Evidence estimation} The TI algorithm misses the mark by \zdiff=$11.53\%$, with the worst overall likelihood \zlogl=$-0.643$. TI+ improves this to \zdiff=$1.43\%$, with a likelihood ratio $\exp(2.658-(-0.643))=26.87$. The SS yields a log-likelihood of \zlogl=$2.618$. Its SS+ counterpart marginally reduces \zdiff~from 1.25\% to 1.21\%, with \zlogl=$2.771$. The H+ method outperforms SS+ and TI+, achieving \zlogl=$1.099$ and \zlogl=$3.082$. Out of the DNS methods, dyn-u finds all modes (extremely accurate \zdiff=0.201) albeit at the cost of many likelihood calls. Meanwhile, slice-based samplers are faster but slightly undersampled some modes (evidence biases).
dyn-u has the highest log-likelihood out of DNS methods, at $1.828$. Against H+, it gives a ratio of $\exp(3.082-1.828)=3.5$, almost four times as likely.

\begin{table}
    \centering
    \caption{\label{tab:eggbox_zest} Egg-box evidence SAR estimation comparison.}
        \begin{tabular}{lllll}
            \toprule
                       &  \lzest      & \lzerrest  & \zdiff (\%)& \zlogl\\
            \midrule
            \midrule
            TI         & 235.733±0.013 &0.748±0.008 &11.532  &-0.643  \\
            SS         & 235.843±0.012 &0.008±0.001 &\textbf{1.252}   &\textbf{2.618}  \\
            H          & 235.802±0.013 &0.130±0.006 &5.283   &1.036  \\
            \midrule
            TI+        & 235.842±0.013 &0.023±0.008 &1.425   &2.648  \\
            SS+        & 235.844±0.013 &0.008±0.001 &1.211   &2.771  \\
            H+         & 235.845±0.013 &0.010±0.003 &\textbf{1.099}   &\textbf{3.082}  \\
            \midrule
            dyn-u       & 235.858±0.104&    0.064±0.001&    \textbf{0.201} & \textbf{1.828} \\
            dyn-s       & 235.903±0.104&    0.064±0.001&    4.812   & 1.560 \\
            dyn-rs      & 235.953±0.118&    0.064±0.001&    10.186   & 0.679 \\
            \bottomrule
        \end{tabular}
        
    \tablefoot{\reddemc's adaptive algorithms compared to \dynesty's sampling methods. From left to right, the log-evidence estimate, the estimate error, the difference to the true value $\ln{\mathcal{Z}}=235.856$ in percentage \zdiff, and the log-likelihood of the estimator \zlogl.}
\end{table}

\subsection{Hybrid Rosenbrock function}\label{sec:benchmarks_banana}
\begin{figure}
    \includegraphics[width=\columnwidth]{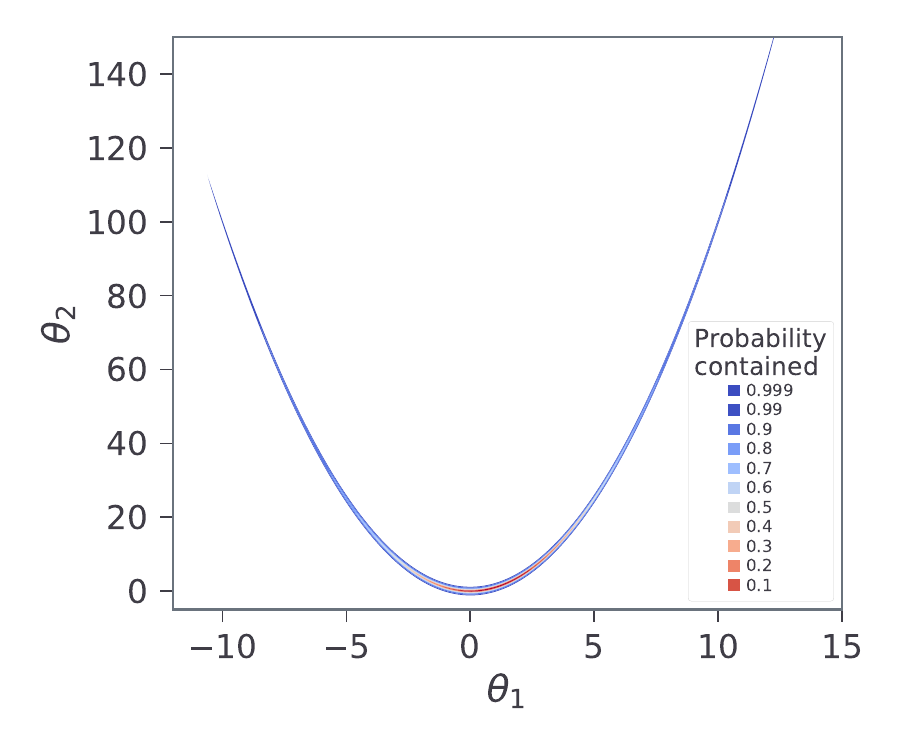}
    \caption{\label{fig:banana_2d}Probability contour of the 2-d Rosenbrock function.  The inset key shows how the colours relate to the probability.}
\end{figure}

The Rosenbrock function is a long, thin, curved valley with strong non-linear correlations, slightly resembling a banana. The hybrid-Rosenbrock function is an extension that provides analytic evidence for any dimension \citep{pagani19_rosenbrock}.
Its likelihood is
\begin{equation} \label{eq:banana}
        p(\vec{x}) = -a(x_1-\mu)^2 - \sum_{j=1}^{n_2}\sum_{i=2}^{n_1}b_{j, i}(x_{j, i}-x_{j, i-1}^2)^2\,,
\end{equation}
where $x_{j,i}, \mu \in \mathbb{R}$, and $a$, $b_{j,i}\in\mathbb{R}^+$ are arbitrary constants. And its evidence is
\begin{equation} \label{eq:banana_evidence}
    \mathcal{Z} = \sqrt{\frac{\pi^n}{a\cdot \prod b_{j,i}}}\,,
\end{equation}
where $n = (n_1-1)\cdot n_2 + 1$, and $n_1$ is the number of dimensions in each of the $n_2$ groups. Per \cite{goodman10}, a scale factor of $1/20$ is chosen, with $a=1/20$ and all $b_{j,i}=100/20$, so the distribution is shaped like a narrow ridge, providing a challenging posterior shape (see \hyperref[fig:banana_2d]{Fig. \ref{fig:banana_2d}}).

To assess temperature evolution in the 2-d case, the APT setup was changed to $[24, 120, 1024, 1]$, increasing the number of temperatures to 24 and the sweeps to 1\,024, while decreasing the walkers to 120, roughly maintaining the total iterations the same as in previous benchmarks.

\paragraph{Sampler performance} The SAR method ranks best with $14.65\pm0.59$ kenits, followed closely by SGG, GAO, and ETL. SMD ranks last, with $6.40\pm0.41$ kenits. For the DNS methods, dyn-rs ranks first with $0.77\pm0.02$ kenits, and dyn-u ranks last with $0.34\pm0.02$ kenits, and almost twice the wall-time (see \hyperref[tab:banana_enits]{Table \ref{tab:banana_enits}}). In this scenario, SAR presents increased kenits over dyn-rs by a factor of $19.03$, almost 20 times more efficient at producing independent samples.

\begin{table}
    \centering
    \caption{\label{tab:banana_enits}2-d Hybrid Rosenbrock performance benchmark.}
        \begin{tabular}{llll}
            \toprule
            Method          & Time (s) & Eff (\%)   & Kenits\\
            \midrule
            \midrule
            SAR         & 19.15±0.20&   9.51±0.38&   \textbf{14.65±0.59} \\
            SMD         & 19.14±0.14&   4.15±0.28&   6.40±0.41  \\
            SGG         & 19.44±0.26&   9.03±0.53&   13.70±0.69 \\
            GAO         & 19.41±0.13&   8.89±0.48&   13.51±0.74 \\
            ETL          & 19.59±0.18&   9.04±0.50&   13.61±0.83  \\
            \midrule
            dyn-u       & 38.28±2.85&   8.04±0.81&   0.34±0.02 \\
            dyn-s       & 21.74±0.93&   2.18±0.06&   0.59±0.03 \\
            dyn-rs      & 16.52±0.42&   2.72±0.09&   0.77±0.02 \\
            \bottomrule
        \end{tabular}
        
    \tablefoot{\reddemc's adaptive algorithms compared to \dynesty's sampling methods with the 2-d Rosenbrock function. From left to right: Time--run time in seconds, Eff--sampling efficiency or percentage of independent samples, and kenits--thousand effective samples per second.}
\end{table}

\paragraph{Evidence estimation} The H+ method relies completely on SS+, dropping TI+ segments (see \hyperref[tab:banana_zest]{Table \ref{tab:banana_zest}}). This can be attested by the poor TI+ performance compared to SS+. The slice-sampling methods present negative log-likelihoods, indicating substantially underestimated errors.

A 3-d case for the Rosenbrock function was also explored (results discussed in Section \ref{sec:discussion}); generally the trends held, with \reddemc~outperforming in sampling efficiency, and all ladder methods producing accurate evidences (see Table \ref{tab:banana3_zest}).

\begin{table}
    \centering
    \caption{\label{tab:banana_zest} 2-d Hybrid Rosenbrock SAR evidence-estimation comparison.}
        \begin{tabular}{lllll}
            \toprule
            Method      &  \lzest      & \lzerrest  & \zdiff (\%)& $\ln\mathcal{L}_{\hat{\mathcal{Z}}}$\\
            \midrule
            \midrule
            TI  & 1.416±0.022 &0.912±0.025 &34.431  &-0.934   \\
            SS  & 1.833±0.020 &0.011±0.001 &\textbf{0.515}   &\textbf{3.479}   \\
            H   & 1.833±0.020 &0.011±0.001 &\textbf{0.515}   &\textbf{3.479}   \\
            \midrule
            TI+ & 1.781±0.022 &0.227±0.017 &5.565   &0.531   \\
            SS+ & 1.833±0.022 &0.012±0.001 &\textbf{0.515}   &\textbf{3.399}   \\
            H+  & 1.833±0.022 &0.012±0.001 &\textbf{0.515}   &\textbf{3.399}   \\
            \midrule
            
            dyn-u  & 1.881±0.089&    0.076±0.001& \textbf{4.394}    & \textbf{1.501}      \\
            dyn-s  & 2.007±0.221&    0.074±0.004& 18.412   & -0.936     \\
            dyn-rs & 1.979±0.164&    0.074±0.002& 15.258   & -0.144     \\
            \bottomrule
        \end{tabular}
        
    \tablefoot{\reddemc's adaptive algorithms compared to \dynesty's uniform (dyn-u), slice (dyn-s), and random-slice (dyn-rs) sampling methods. From left to right, the log-evidence estimate, the estimate uncertainty, the difference to the true value $\ln{\mathcal{Z}}$=$1.838$ in percentage \zdiff, and the log-likelihood of the estimator \zlogl.}

\end{table}

\section{Exoplanet detection from radial velocities} \label{sec:hd20794}

A real-world application of the APT ladder dynamics developed in Section \ref{sec:temp_ladder} \reddtext{and} evidence-estimation methods introduced in Section \ref{sec:evidence_error_est}, is presented for the nearby G-dwarf HD\,20794. 
Hosting at least three super-Earths whose RV semi-amplitudes are lower than 1 $ms^{-1}$, the system combines low amplitudes, multiplicity, and a long-period planet, all features that make model selection and evidence estimation particularly demanding. The innermost signals at 18 and 90 days were first identified in the HARPS discovery paper \citep{HD20794_pepe11}, while subsequent analyses revealed an outer candidate with a period of $\sim$650 d that spends a significant fraction of its eccentric orbit inside the stellar habitable zone \citep{HD20794_nari25}.

As such, this system provides an ideal test-bed for assessing the performance of our APT dynamics and evidence-estimation algorithms in a real-world example. \reddtext{The main features that make exoplanet RV fitting challenging are directly reflected in the previous benchmarks: the strong phase transition in $C_{\nu}$ is captured by the Gaussian shells, multimodality and mode completeness in a many-alias situation (for a sparsely sampled period) in the Gaussian egg-box, and the curved, highly-correlated ridges typically seen in the angular orbital parameters are emulated by the hybrid Rosenbrock function}. In this section, the work by \cite{HD20794_nari25} is followed, applying \reddemc~to the combined HARPS+ESPRESSO time-series, contrasting the resulting evidences with those reported in the literature.

\subsection{Model and likelihood}

Each exoplanet is characterised by five Keplerian parameters, $P$--the period, $K$--the semi-amplitude, $e$--the eccentricity, $\bar{\omega}$--the longitude of periastron, and $M_0$--phase of periastron passage:

\begin{equation} \label{eq:keplerian}
    \mathcal{K}_j(t) = K_j \cdot [\cos(\nu_j(t, P_j, e_j, M_{0j}) + \bar{\omega}_j) + e_j\cos(\bar{\omega}_j)]\,,
\end{equation}

where the subscript $j$ indicates each exoplanet and $\nu_j$ is the true anomaly.
In addition, each instrument is given an offset $\gamma$ (an additive constant) and a jitter $\sigma_{\mathrm{INS}}$ (added in quadrature to the measurement error, corresponding to a white noise component). Therefore, under the assumption of Gaussian-like errors, the log-likelihood is

\begin{equation}
    \ln{\mathcal{L}} = -\frac{1}{2}\sum_{\mathrm{INS}}\sum_{i}^{N_{\mathrm{INS}}}\left(\frac{\xi^{2}_{i,\mathrm{INS}}}{(\sigma^{2}_i + \sigma_{\mathrm{INS}}^{2})} + \ln(\sigma^{2}_i + \sigma_{\mathrm{INS}}^{2})\right) - \frac{N\ln(2\pi)}{2}\,,
\end{equation}

where the sub-indices $i$ and $\mathrm{INS}$ correspond to the $i$-th RV measurement (taken at a time $t_i$) and to each instrument, respectively. The residuals $\xi_{i,\mathrm{INS}}$ are defined as the difference between the data and the model. Three distinct datasets plus three Keplerian signals add up to a total of 21 parameters or dimensions.

\subsection{Parameter estimation and model selection}
For an in-depth methodology follow-up, refer to \refapp{sec:appendix_hd20794}.
The models compared are: 1) just white noise ($H_0$); 2) a single sinusoid ($1S$); 3) two sinusoids ($2S$); 4) three sinusoids ($3S$); 5) 3 Keplerians ($3K$); and 6) 4 Keplerians ($4K$). The sinusoidal models (S) treat planets as having circular orbits. The $3K$ model corresponds to the three confirmed planets, and $4K$ to the inclusion of an additional candidate.

For the evidence estimation (see \reftab{tab:hd20794}), dyn-rs was consistent with SAR, up to the most complex models. For the $3S$, $3K$, and $4K$ models, they had a 4.6, 6.8, and 4.6 $\ln\mathcal{\widehat{Z}}$ difference, respectively. On the other hand, the SAR log-evidences were all within the reference values' uncertainties except for the $1S$ case, which showed a moderate discrepancy ($\sim$8.7). This could be due to random variation or differences in how jitter/noise was treated, but importantly the model ranking is unaffected.

\reftab{tab:hd20794_eff} compares the estimated evidence errors to the actual run-to-run scatter (the ratio $\widehat{\sigma}_{\mathcal{Z},\mathrm{estimated}}/\widehat{\sigma}_{\mathcal{Z},\mathrm{empirical}}$). An ideal ratio is 1. Values $\ll$1 mean the method underestimates its true uncertainty, and $\gg$1 means overestimation. The dyn-rs ratios are all very low for complex models (0.04-0.12), severely underestimating the uncertainty with overly confident evidences, whereas \reddemc~yields values much closer to unity (e.g., 0.63 for the $3K$). Underestimation of error in multi-planet cases could lead to overconfident claims of detection, therefore, \reddemc~(along the H+ estimator) provides more reliable uncertainties.

Performance-wise, APT achieves a higher kenits count in the simpler models, with a turnover point at $3K$ in favour of dyn-rs. This may be attributable to the modest fixed number of walkers (256) in the APT sampler (see \refapp{sec:appendix_hd20794} for details). In practice, one could adjust walker count for efficiency's sake, but it was kept constant for consistency \reddtext{between benchmarks}.

\begin{table}
    \centering
    \caption{\label{tab:hd20794} HD\,20794 evidence-estimation comparison.}
        \begin{tabular}{llll}
            \toprule
            Model  &    Reference$^{(1)}$   & dyn-rs& SAR \\
            \midrule
            \midrule
            $H_0$ &-111.9±0.2 &-111.2±0.1 &-111.9±0.1 \\
            1 S   &-72.9±3.0  &-65.1±2.7  & -64.2±0.1 \\
            2 S   &-41.7±3.7  &-40.2±2.4  & -40.2±0.2 \\
            3 S   &-12.3±2.0  &-16.0±1.1  & -11.5±0.1 \\
            3 K   &-3.6±0.0$^{(2)}$ &-12.9±2.6&-5.8±0.5\\
            4 K   &-1.6±4.6 &-4.6±3.8     &0.0±1.0    \\
            \bottomrule
        \end{tabular}
    \tablebib{(1)~\citet{HD20794_nari25}; (2) unreported standard deviation.}
    \tablefoot{Difference of the model's mean evidence with the evidence of the best-ranking model. The models shown from top to bottom are, white noise only ($H_0$), increasing sinusoids (S), and increasing Keplerians (K). The reference evidences have an offset added to make the $H_0$ evidence equivalent to the SAR.}

\end{table}

\begin{table}
    \centering
    \caption{\label{tab:hd20794_eff} HD\,20794 evidence-estimation uncertainty and performance.}
        \begin{tabular}{lllll}
            \toprule
            & \multicolumn{2}{c}{$\frac{\widehat{\sigma}_{\ln{\widehat{\mathcal{Z}}}}}{\sqrt{\mathrm{Var}[\ln{\mathcal{\widehat{Z}}}]}}$} & \multicolumn{2}{c}{kenits (eff)} \\
            Model  & dyn-rs& SAR & dyn-rs& SAR      \\
            \midrule
            \midrule
            $H_0$ & 2.30 & 11.4 & 0.375 (2.5)  & 0.889 (7.8) \\
            1 S   & 0.05 & 0.48 & 0.193 (1.3)  & 0.679 (4.7)  \\
            2 S   & 0.07 & 0.75 & 0.120 (1.0)  & 0.242 (1.9) \\
            3 S   & 0.12 & 2.03 & 0.097 (0.8)  & 0.207 (1.7) \\
            3 K   & 0.07 & 0.63 & 0.051 (0.5)  & 0.046 (0.4) \\
            4 K   & 0.04 & 0.27 & 0.039 (0.5)  & 0.030 (0.3)  \\
            \bottomrule
        \end{tabular}
    \tablefoot{Left columns show the ratio of the mean of the estimated evidence uncertainty against the standard deviation of the run-to-run evidence. Right columns, the average kenits, with the efficiency in parenthesis. The models are white noise only ($H_0$), increasing sinusoids (S), and increasing Keplerians (K).}

\end{table}

For the parameter estimation, both methods produce consistent results with each other (see \reftab{tab:hd20794_param}). Notably, the APT-derived parameter posteriors are slightly more asymmetrical and tighter (see $K_2$ and $K_3$). Also, the eccentricity estimates seem less averse to boundary solutions, as seen in $e_1$=$0.057^{+0.045}_{-0.057}$ and $e_2$=$0.026^{+0.007}_{-0.026}$, both consistent with $e=0$.

\begin{table}
    \centering
    \caption{\label{tab:hd20794_param} HD\,20794 parameter estimation.}
        \begin{tabular}{llll}
            \toprule
            Parameter  &    Reference$^1$   & dyn-rs& SAR \\
            \midrule
            \midrule
            $P_1$ (days) & 18.314±0.002             & $18.314^{+0.004}_{-0.004}$& $18.313^{+0.001}_{-0.001}$  \\
            $K_1$ (ms$^{-1}$)& 0.614±0.048              & $0.632^{+0.030}_{-0.066}$ & $0.608^{+0.030}_{-0.009}$  \\
            $e_1$ & $0.064^{+0.065}_{-0.046}$& $0.109^{+0.049}_{-0.068}$ & $0.057^{+0.045}_{-0.057}$  \\
            \midrule
            $P_2$ (days) & 89.68±0.10               & $89.67^{+0.11}_{-0.16}$   & $89.73^{+0.06}_{-0.03}$     \\
            $K_2$ (ms$^{-1}$) & $0.502^{+0.048}_{-0.049}$& $0.510_{-0.054}^{+0.038}$ & $0.474^{+0.041}_{-0.001}$  \\
            $e_2$ & $0.077^{+0.084}_{-0.055}$& $0.055_{-0.054}^{+0.053}$ & $0.026^{+0.007}_{-0.026}$  \\
            \midrule
            $P_3$ (days) & $647.6^{+2.5}_{-2.7}$    & $652.1^{+4.5}_{-5.5}$     & $648.2^{+2.6}_{-2.4}$       \\
            $K_3$ (ms$^{-1}$) & $0.567^{+0.067}_{-0.064}$& $0.521_{-0.091}^{+0.025}$ & $0.589^{+0.001}_{-0.055}$  \\
            $e_3$ & $0.45^{+0.11}_{-0.10}$   & $0.428_{-0.412}^{+0.111}$ & $0.467^{+0.053}_{-0.024}$  \\

            \bottomrule
        \end{tabular}
    \tablebib{(1)~\citet{HD20794_nari25}.}
    \tablefoot{Parameter estimation results for the different algorithms for the $3 K$ model. In descending order, $P$ corresponds to the period, $K$ to the semi-amplitude, and $e$ to the eccentricity, where the subscript denotes a particular planet.}

\end{table}

\section{Discussion} \label{sec:discussion}

The evidence-estimation algorithms introduced in Section \ref{sec:evidence_error_est} have proven to be far more reliable than their unmodified counterparts. Furthermore, whether to apply the TI or the SS method is mostly problem-dependent, and the hybrid algorithm seems to be a good compromise when there is no a priori information about the problem at hand.  Moreover, our proposed evidence estimators provide a solid alternative, if not better, to dedicated evidence-sampling methods. This is exemplified by the real-world case studied in Section \ref{sec:hd20794}.

Across the simplest problems we tested (few dimensions, nearly constant specific heat), all ladder strategies converge to an almost geometric ladder and yield similar performance. Differences emerge as the problem's complexity grows, and since a common problem astrophysicists face is that of dimensionality, we further discuss it next.

\subsection{Increasing dimensionality} \label{sec:discussion:increasing_dimension}

We take the Gaussian shells problem described in \refsec{sec:benchmarks_gaussian_shells} and increase its dimensionality without modifying any hyper-parameters, in order to demonstrate how the methods scale. As dimensionality grows, we see a very much expected efficiency decrease (see \reftab{tab:shells_dim}), as well as the SMD outperforming other methods. At 15 dimensions, it becomes the most effective, with a margin far exceeding the run-to-run scatter. In second place, GAO and ETL come tied (with overlapping confidence intervals), being around 14\% slower, followed by the SAR--at 19\% slower--and the SGG--being 34\% slower. This supports the intuition that maximising swap distance is beneficial in high-dimensional spaces where local energy traps abound. On the other hand, dyn-rs, still the best DNS method, presents kenits at $\sim$13\% of the SMD average.

\begin{table}
    \centering
    \setlength{\tabcolsep}{5pt}
    \caption{\label{tab:shells_dim}Gaussian shells with increasing dimensions performance.}
        \begin{tabular}{lllll}
            \toprule
            Method      & 2-d         & 5-d        & 10-d       & 15-d        \\
            \midrule
            \midrule
            SAR         &   3.70±0.12 &3.59±0.12   &2.71±0.11   & 2.25±0.07   \\
            SMD         &   3.73±0.13 &3.73±0.14   &3.21±0.14   & \textbf{2.80±0.09}   \\
            SGG         &   3.67±0.13 &3.50±0.11   &2.45±0.07   & 1.85±0.07   \\
            GAO         &   3.60±0.11 &3.61±0.15   &2.84±0.07   & 2.36±0.07   \\
            ETL          &   3.66±0.10 &3.66±0.12   &2.89±0.07   & 2.45±0.13   \\
            \midrule
            dyn-u       &   0.19±0.04 &0.37±0.04   &0.32±0.05   & 0.18±0.01   \\
            dyn-s       &   0.75±0.04 &0.43±0.01   &0.23±0.01   & 0.17±0.01   \\
            dyn-rs      &   0.94±0.05 &0.70±0.01   &0.46±0.01   & 0.37±0.01   \\

            \bottomrule
        \end{tabular}
        
    \tablefoot{\reddemc's adaptive algorithms compared to \dynesty's sampling methods. The kenits--effective samples per second in thousands--are shown from left to right with increasing dimensions.}
\end{table}

For evidence estimation, with the modest amount of temperatures proposed for this benchmark, the TI method degrades considerably, with a \zdiff~of $\sim72\%$ (see \reftab{tab:15dgs_zest}). The TI+ soothes this, with a \zdiff~$11.3\%$. Nevertheless, the error due the discretisation remains a problem in this benchmark. For the SS method, the evidence estimate is excellent (\zdiff=$2.56$), with the highest \zlogl=$2.1658$. The SS+ method presents a slight deterioration over its classic counterpart. The H+ presents a higher likelihood (and lower \zdiff) than either TI+ or SS+.

By increasing the dimensionality of the Rosenbrock function to 3-d, the evidence estimation results (H+) are best for ETL with \zlogl=$2.901$, followed by SGG and GAO, with \zlogl=1.841 and 1.561, respectively (see \reftab{tab:banana3_zest}). The kenits values present a similar trend to that of the shells, with 2.98 (SAR), 2.50 (SMD), 3.46 (SGG), 3.03 (GAO), and 2.85 (ETL); and 0.02 (dyn-u), 0.55 (dyn-s), and 0.74 (dyn-rs).

\begin{table}
    \centering
    \caption{\label{tab:banana3_zest} Hybrid 3-d Rosenbrock evidence estimation.}
        \begin{tabular}{lllll}
            \toprule
            Method      &  \lzest      & \lzerrest   & \zdiff& $\ln\mathcal{L}_{\hat{\mathcal{Z}}}$\\
            \midrule
            \midrule
            SAR  &1.562±0.062 &0.021±0.001 &4.233   &0.771 \\
            SMD  &1.516±0.065 &0.023±0.002 &8.519   &-4.556 \\
            SGG  &1.574±0.076 &0.021±0.002 &3.069   &1.841 \\
            GAO  &1.570±0.042 &0.022±0.002 &3.500   &1.561  \\
            ETL   &1.598±0.061 &0.021±0.001 &\textbf{0.747}   &\textbf{2.901} \\

            \midrule
            dyn-u  & 1.923±0.207 & 0.109±0.002& 37.366  &-2.944  \\
            dyn-s  & 1.625±0.883 & 0.111±0.004& 1.967   & 1.264  \\
            dyn-rs & 1.534±0.442 & 0.112±0.006& 6.903   & 1.066  \\
            \bottomrule
        \end{tabular}
        
    \tablefoot{\reddemc's adaptive algorithms compared to \dynesty's sampling methods. From left to right, the log-evidence estimate, the estimate error, the difference to the true value $\ln{\mathcal{Z}}$=$1.606$ in percentage in percentage \zdiff, and the log-likelihood of the estimator \zlogl.}

\end{table}

\subsection{Burn-in}
How fast the adaptation optimally distributes the ladder depends on both the problem at hand and the adaptation hyper-parameters $\nu_0$ and $\tau_0$. We stress test adaptation by setting the 15-d shells with 640 sweeps (as in \refsec{sec:benchmarks_eggshells} and \refsec{sec:discussion:increasing_dimension}) to use 320, 160, 64, and 32 sweeps as burn-in (see \reftab{tab:15dgs_zest_burn}). H+ is the method most resilient to an uncalibrated ladder, it is the only method that retains positive \zlogl~values after the 25\% burn-in, with values $\ln\widehat{\mathcal{Z}}|_{H+}$=-24.939±0.037, -24.980±0.031, -24.984±0.034, and -24.996±0.041, for 50\%, 25\%, 10\%, and 5\%, respectively.

\subsection{Ladder adaptation}
For low-dimensional targets (2-d shells, egg-box) every method settles on an almost geometric ladder, due to $C_{\nu}(\beta)$ being almost flat. As the dimensionality grows, $C_{\nu}$ develops a broad peak around the phase-transition region, and the objectives react differently.
For example, the SAR spreads the temperatures uniformly in swap probability and therefore stretches the ladder on both sides of the peak, while SMD concentrates the replicas where the specific heat is large, maximising the average information transfer, as seen by its clear lead at 15-d in \reftab{tab:shells_dim}.

Longer runs are realised for this scenario, 15-d shells, with 10\,000 sweeps instead of 640, to examine the convergence of the ladder adaptation. All schemes compress the first $\sim$200 sweeps into a fast logistic-like relaxation of the temperature gaps, after which the diminishing adaptive factor $\kappa(t)$ (\refeq{eq:kappa}) forces a slow power-law tail. In practice, almost the whole ladder shape is set in under 5\% of the computational budget, supporting the choice of freezing the ladder after a user-defined burn-in.

\begin{figure}
\centering
    \includegraphics[width=\columnwidth]{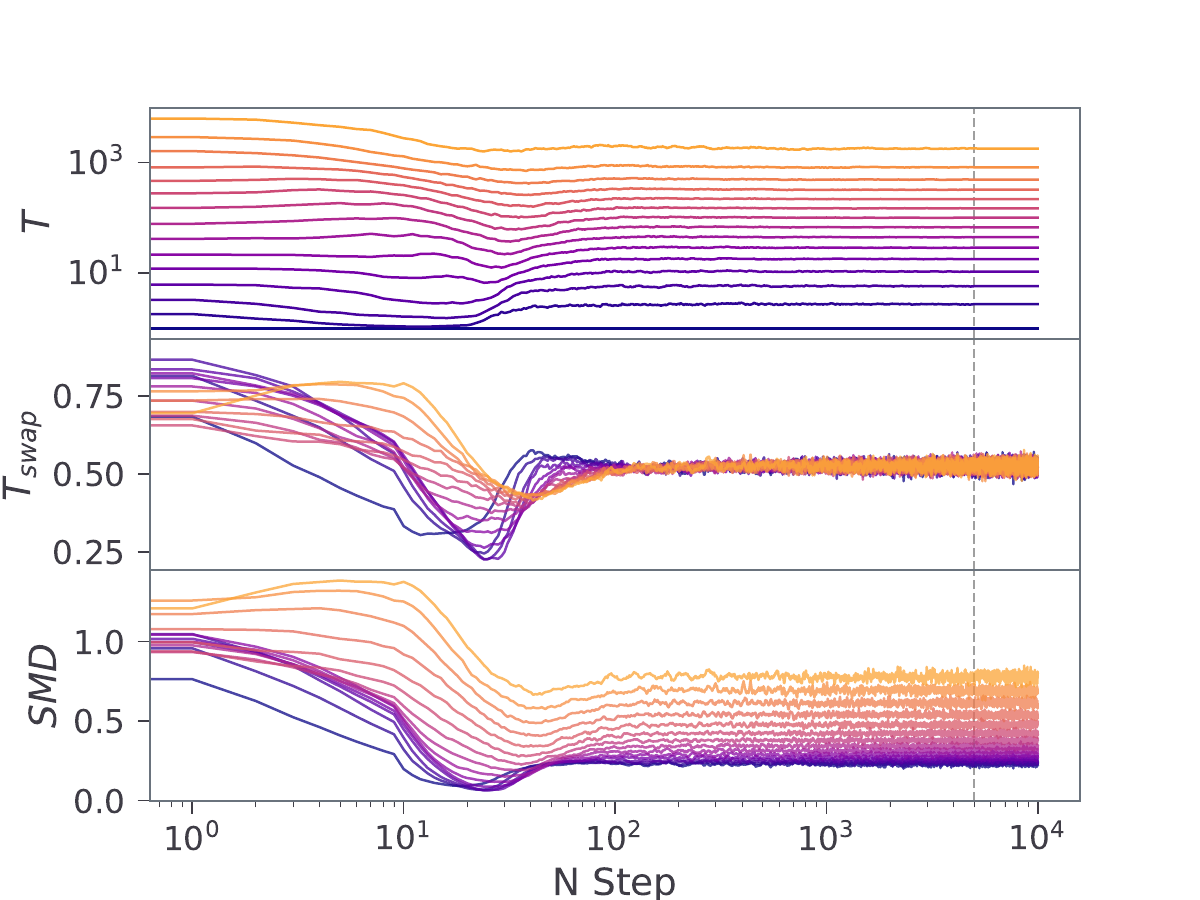}
    \includegraphics[width=\columnwidth]{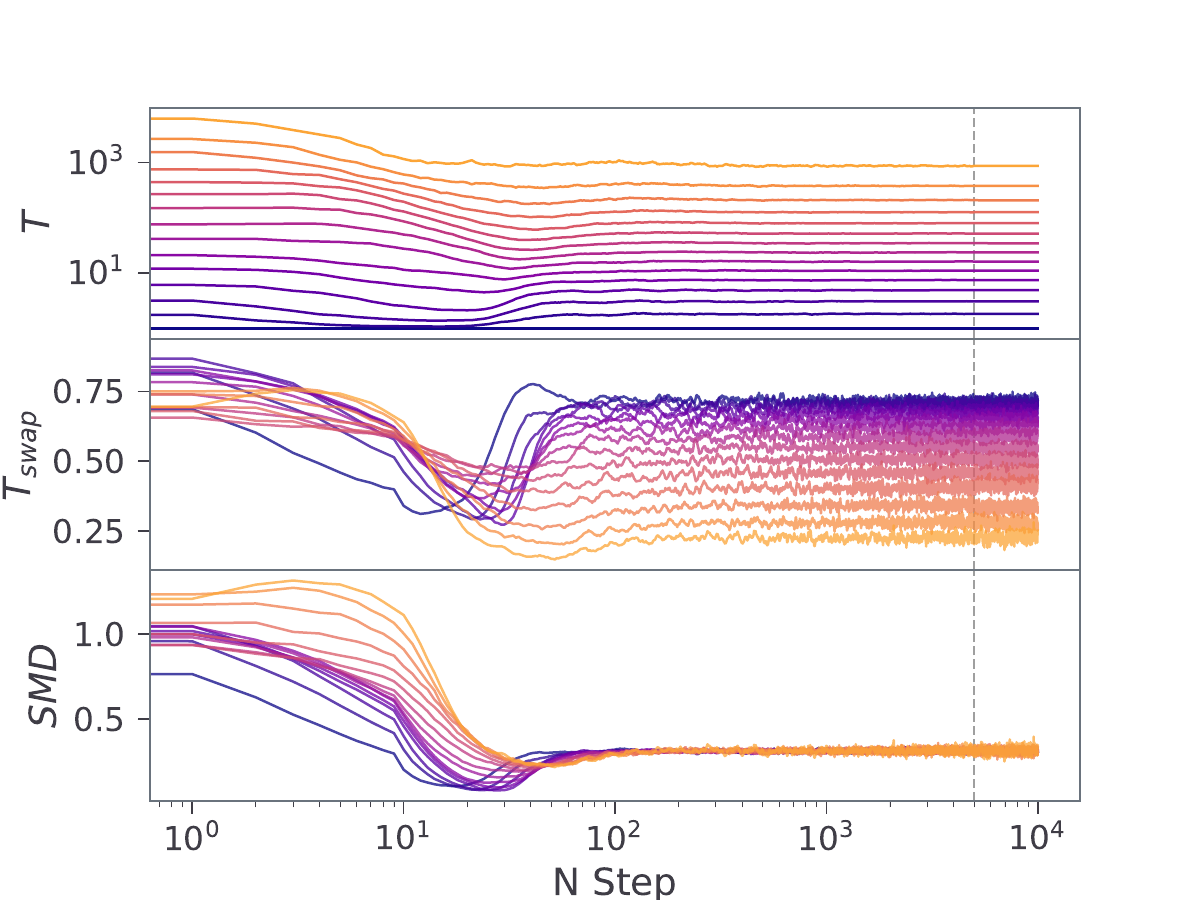}\label{fig:shells_15_ladder}
        \caption{Temperature ladder evolution for the 15-d Gaussian shells in the SAR regime (top) and SMD regime (bottom). In the x-axis is the current iteration. Descending, the temperature ladder evolution $T$, the swap acceptance ratio $T_{swap}$, and the swap mean distance SMD. Colours represent each chain, where blue is the coldest ($\beta$=1), with increasing temperature to the red, where the hottest chain is omitted. The vertical black dashed line indicates where the adaptation stops.}
\end{figure}

\reffig{fig:shells_15_ladder} displays the ladder evolution of the SAR and SMD methods, respectively. A notable difference is that the SMD method has a layered swap rate, where the cold chain swaps frequently (0.7), and the hot chain seldom (0.2), whereas the SAR keeps a steady swap rate of 0.5 for all temperatures.

\begin{figure}
    \includegraphics[width=\columnwidth]{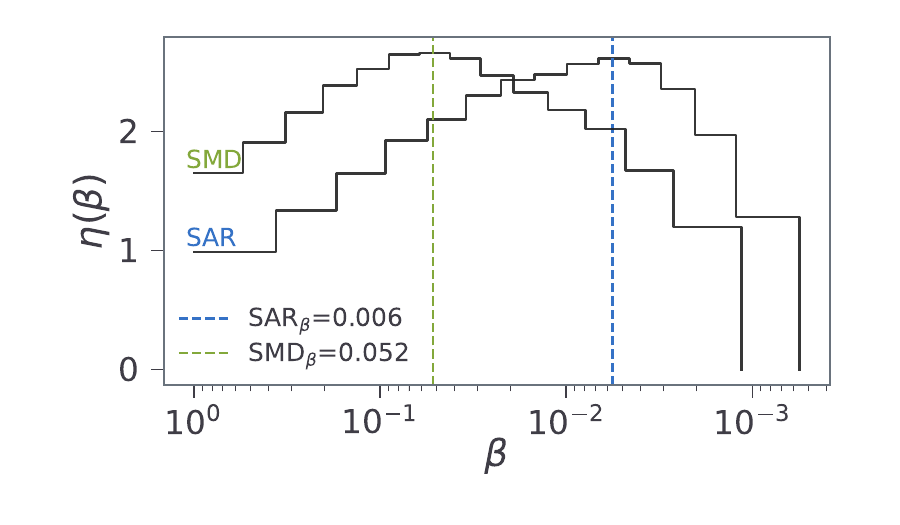}
    \caption{\label{fig:shells_15_density}Chain density per temperature range for the SAR and SMD methods in the 15-d Gaussian shells. The blue dashed line denotes the maximum of the SAR, with the green one representing the maximum of the SMD.  Both lines also denote the regions where the TI or SS algorithms are applied in the Hybrid method.}
\end{figure}

\reffig{fig:shells_15_density} shows the final ladders as chain density (proportional to $C_{\nu}$) as a function of temperature. The dashed vertical lines indicate where $\Delta\beta$ is smaller, as well as where the transition between TI and SS is applied in the hybrid method.

\subsection{Number of temperatures}
\begin{figure}
    \includegraphics[width=\columnwidth]{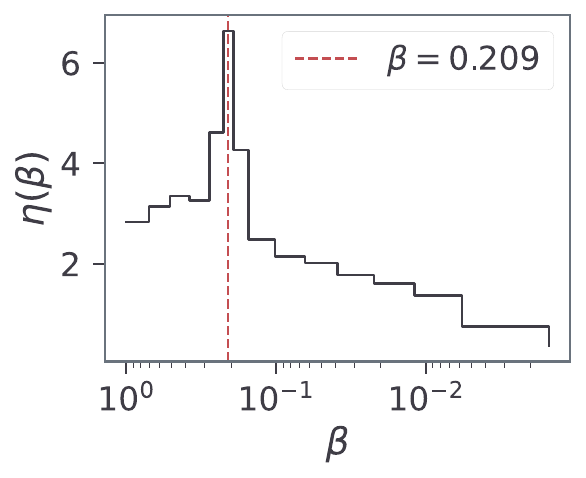}
    \caption{\label{fig:hd20794_density}Chain density per temperature range for the HD20794 $3K$ SAR run.  The vertical dashed red line marks the distribution maximum.}
\end{figure}

A natural question that arises is `how many temperatures are enough?' The answer, as one might expect, is `it depends'. When doing an APT run, there are two objectives in mind, posterior and evidence estimation. More temperatures might be redundant posterior wise, not contributing with additional information to the cold chain. At the same time, more temperatures provide a finer grid for the evidence estimation.
Nonetheless, it is possible to dabble into this question, by analysing the temperature evolution of the ladder. If the mixing is low, increasing the temperatures will result in a more efficient run. This may be measured by comparing the additional chains computational cost against the efficiency (or kenits) increase.
Visualising the likelihood variance as a function of temperature (the thermodynamic integration) can be helpful. A smoothly increasing curve will not be affected as much as a ladder-like curve by decreasing temperatures. This can also be appreciated by the $C_{\nu}$ plot, a ladder-like behaviour would be seen as multiple sharp peaks (as in the HD\,20794 $3K$ model, see \reffig{fig:hd20794_density}, with a secondary small peak at around $\beta$=0.4), whereas a smooth curve is seen as a single smooth peak (like in the 15-d shells, see \reffig{fig:shells_15_density}).

But how are the different evidence-estimation algorithms proposed affected? The 15-d shells (with 10\,000 sweeps) is re-run with $N_{\beta}$=6, 8, 12, and 16 (see \reftab{tab:15dgs_zest_temp}). The TI method by decreasing temperatures, presents a linear fall-off with $\sqrt{\mathcal{L_{\widehat{Z}}}}$, this result can also be easily derived from \refeq{eq:zerr_disc}, which will reveal that the log-evidence error is proportional to $\frac{1}{\mathrm{ntemps}}$.
The TI+ method presents a huge improvement over TI, although still failing to reach the mark in temperature poor regimes.
The SS methods remain extremely accurate even at low $N_{\beta}=6$, with \zdiff=1.78 and \zlogl=2.02. The geometrical bridge in SS+ yields a small accuracy gain (lower \zdiff~for every $N_{\beta}$), while bringing a small drop in \zlogl, due the slight increase in the error estimate.

\section{Conclusions}  \label{sec:conclusions}

We have introduced \reddemc, an adaptive parallel tempering ensemble sampler that couples three next-level techniques--flexible ladder adaptation, robust evidence estimation, and practical error quantification--into a single, easy-to-deploy package. Three of the five ladder adaptation algorithms are new implementations from this work (SMD, SGG, ETL), reliable alternatives to the most commonly used SAR method, as demonstrated in our benchmarks (see Section \ref{sec:benchmarks}). All of these methods converge in a few sweeps and deliver cold-chain mixing efficiencies around an order of magnitude higher than dynamic nested sampling across our tests. In the challenging 15-d Gaussian-shell benchmark, \reddemc~sustained $>2$ kenits, about 7 times faster than the best DNS configuration, while the SMD ladder proved particularly resilient as dimensionality grew.

We devised three evidence estimators--two original implementations, TI+ and SS+, and a novel hybrid approach--that combine curvature-aware interpolation with bridge sampling. These estimators yield accurate log-evidences and maintain realistic uncertainties even when the number of temperatures is decreased beyond optimal. 

A real-world application to the HD\,20794 radial-velocity dataset shows that \reddemc~ reproduces literature model rankings, recovers planetary parameters with tighter--yet statistically consistent--credible intervals, and supplies evidence uncertainty that closely tracks run-to-run dispersion. Crucially, this performance is obtained without manual ladder tuning: the same hyper-parameters were used from no-planet to four-planet models and from 6 to 21 dimensions.

Overall, \reddemc~demonstrates that a carefully engineered APT sampler can match, and occasionally surpass, state-of-the-art DNS, both in sampling throughput and in evidence estimation, while retaining the posterior-inference strengths that make MCMC indispensable. Future work will explore on-the-fly convergence diagnostics, further widening the sampler's applicability to the increasingly complex problems faced in modern astrophysics and beyond.

\begin{acknowledgements}
    PAPR and JSJ gratefully acknowledge support by FONDECYT grant 1240738, from the ANID BASAL project FB210003, and from the CASSACA China-Chile Joint Research Fund through grant CCJRF2205.
    For the n-d Gaussian shells, Gaussian egg-box, and Rosenbrock function benchmarks parallelisation was not used and computing was limited to single-core for all algorithms. For the exoplanet detection benchmark, parallelisation was used with 24 threads. All the benchmarks were performed on a computer with an AMD Ryzen Threadripper 3990X 64-Core Processor with 128Gb of DDR4 3200Mhz RAM. We thank the anonymous referee for insightful suggestions that enhanced the quality of this manuscript. We are grateful to Fabo Feng and Mikko Tuomi for stimulating early discussions on planet detection. We would also like to thank Dan Foreman-Mackey for the excellent library \texttt{emcee} \citep{sampler_emcee}, which opened a most exciting new world.
    Typesetting was carried out in \texttt{Overleaf} \citep{Overleaf2025}.
\end{acknowledgements}

\bibliographystyle{aa}
\bibliography{aa56609-25}

\begin{appendix}

\onecolumn
\section{Gaussian shells}\label{sec:appendix_gaussian_shells}

\begin{table*}[h!]
    \centering
    \caption{\label{tab:15dgs_zest} 15-d Gaussian shells evidence-estimation comparison with the SAR method.}
        \begin{tabular}{lllll}
            \toprule
            Method      &  \lzest      & \lzerrest    & \zdiff& \zlogl \\
            \midrule
            \midrule
            TI          &-26.1818±0.0417 &1.1037±0.0526 &71.9266  &-1.6800  \\
            SS          &-24.9374±0.0410 &0.0205±0.0029 &\textbf{2.5617}   &\textbf{2.1658}   \\
            Hybrid      &-24.9523±0.0423 &2.9866±0.0661 &4.0070   &-2.0132  \\
            \midrule
            TI+         &-25.0308±0.0375 &0.5979±0.0266 &11.2557  &-0.4245  \\
            SS+         &-24.9411±0.0369 &0.0204±0.0021 &2.9299   &1.9090   \\
            Hybrid+     &-24.9394±0.0373 &0.0242±0.0027 &\textbf{2.7599}   &\textbf{2.1337}   \\
            \midrule
            dyn-u       &-24.7176±0.3144& 0.1246±0.0017& 21.3853& -0.0460\\
            dyn-s       &-24.8765±0.3044& 0.1243±0.0015& \textbf{3.5620}&\textbf{1.1267} \\
            dyn-rs      &-24.7292±0.2163& 0.1242±0.0012& 19.9854& 0.0908 \\
            \bottomrule
        \end{tabular}
        
    \tablefoot{\reddemc's adaptive algorithms compared to \dynesty's sampling methods. From left to right, the log-evidence estimate, the estimate uncertainty, the difference to the true value $\ln{\mathcal{Z}}$=$-24.9114$ in percentage \zdiff, and the log-likelihood of the estimator \zlogl.}

\end{table*}

\begin{table*}[h!]
    \centering
    \caption{\label{tab:15dgs_zest_burn} 15-d Gaussian shells $\ln\mathcal{Z}$ estimation with variable burn-in for the SAR method.}
        \begin{tabular}{lllllllll}
            \toprule
            Method      & \multicolumn{2}{c}{$N_{\mathrm{adapt}}$=50\%}& \multicolumn{2}{c}{$N_{\mathrm{adapt}}$=25\%}& \multicolumn{2}{c}{$N_{\mathrm{adapt}}$=10\%}   & \multicolumn{2}{c}{$N_{\mathrm{adapt}}$=5\%}   \\
            & \zdiff& \zlogl &
              \zdiff& \zlogl & 
              \zdiff& \zlogl &
              \zdiff& \zlogl \\
            \midrule
            \midrule
            TI        &71.9266  &-1.6800 &72.9348  &-1.6919  &71.0282  &-1.8322  &70.176   &-22.5295  \\
            SS        &2.5617   &2.1658  &7.0111   &-5.9413  &6.7014   &-4.9155  &8.3851   &-8.4817   \\
            Hybrid    &4.0070   &-2.0132 &21.8509  &-1.4656  &27.8067  &-1.0694  &17.2876  &-2.0419   \\
            \midrule
            TI+       &11.2557  &-0.4245 &14.1321  &-0.5039  &12.6212  &-0.3977  &12.3431  &-0.3729   \\
            SS+       &2.9299   &1.9090  &6.3493   &-4.2133  &6.3843   &-4.7487  &8.1264   &-8.5345   \\
            Hybrid+   &2.7599   &2.1337  &6.6707   &1.2514   &6.9972   &1.0890   &8.1322   &0.9562    \\
            \bottomrule
        \end{tabular}
        
    \tablefoot{\reddemc's SAR method evidence estimation with different adaptation (and burn-in) times, starting at half of the chain $N_{\mathrm{adapt}}$=50\%, down to 25\%, 10\%, and 5\% for 640 sweeps. Each sub-column shows the difference to the true value $\ln{\mathcal{Z}}$=$-24.9114$ in percentage \zdiff, and the log-likelihood of the estimator \zlogl.}

\end{table*}

\begin{table*}[h!]
    \centering
    \caption{\label{tab:15dgs_zest_temp} 15-d Gaussian shells $\ln\mathcal{Z}$ estimation with variable temperatures for the SAR method.}
        \begin{tabular}{lllllllll}
            \toprule
            Method      & \multicolumn{2}{c}{$N_{\beta}=16$}& \multicolumn{2}{c}{$N_{\beta}=12$}& \multicolumn{2}{c}{$N_{\beta}=8$}   & \multicolumn{2}{c}{$N_{\beta}=6$}   \\
            & \zdiff& \zlogl &
              \zdiff& \zlogl & 
              \zdiff& \zlogl &
              \zdiff& \zlogl \\
            \midrule
            \midrule
            TI        &70.7557 &-1.6760 &90.4453 &-2.5408 &99.8465 &-4.0657 &>100.0   &-5.2445  \\
            SS        &0.5806  &3.6858  &0.3422  &4.1314  &0.1161  &4.1473  &1.7751   &2.0183   \\
            Hybrid    &15.5426 &-1.8808 &10.6859 &0.0712  &18.2425 &-2.7388 &43.9598  &-3.0542  \\
            \midrule
            TI+       &8.6369  &-0.4128 &25.0197 &-1.3178 &77.9445 &-2.5999 &99.2296  &-3.5185  \\
            SS+       &0.5052  &3.8631  &0.5405  &3.7914  &0.0666  &4.3621  &1.3469   &1.8170   \\
            Hybrid+   &0.6942  &1.8130  &0.5567  &1.7850  &0.8203  &-0.3522 &8.5362   &-1.6289  \\
            \bottomrule
        \end{tabular}
        
    \tablefoot{\reddemc's SAR method evidence estimation with $N_{\beta}$=16, 12, 8, and 6 temperatures. Each sub-column shows the difference to the true value $\ln{\mathcal{Z}}$=$-24.9114$ in percentage \zdiff, and the log-likelihood of the estimator \zlogl.}

\end{table*}

\twocolumn

\section{HD 20794}\label{sec:appendix_hd20794}
Each dataset, HARPS03, HARPS15, and E19, is nightly binned, then sigma clipped ($3\sigma$), and measurements where the error is higher than 3 times the median error are excluded, leaving out 512, 231, and 63 RVs respectively, for a grand total of 806 RVs. Priors were matched to those stated by \cite{HD20794_nari25}, with unspecified priors left to an educated guess. For offset a prior \Uniform{-3}{3} was chosen, and for jitter \Normal{0}{5}, truncated at [0, 3]. For eccentricity $e$ and longitude of periastron $\omega$, we use the change of variable 
\begin{equation} \label{eq:cv-hou}
    e_c = \sqrt{e}\cos(\omega)\,;\quad     e_s = \sqrt{e}\sin(\omega)\,,
\end{equation}
to linearise the circular parameter $\omega$, improving sampler performance, as in \citet{2012ApJ...745..198H}. \reddtext{Point estimates and uncertainties for all parameters are defined as the posterior maxima and the corresponding 1-$\sigma$ highest-density intervals (HDIs), rather than the more commonly used medians and corresponding 1-$\sigma$ percentiles, reflecting our methodological preference for mode-based summaries.}

Both methods had matching estimates for the $H_0$ model (with little variance). This value was chosen as offset for the reference values from the literature.
We tried to follow the recipe of executing five runs and selecting the three with highest evidences. But this could not be applied to all models due to inconsistent results.

For the \dynesty~runs (random-slice sampling, 3\,000 live-points), the $1S$ model consistently found $P_1=18.314$~d, $K=0.61$ ms$^{-1}$, with $\ln\hat{\mathcal{Z}}=-1218.1\pm2.7$, presenting a 46.1 difference against the $H_0$ model $\ln\hat{\mathcal{Z}}=-1264.20\pm0.04$.
The $2S$ run (3\,000 live-points) consistently added a signal with $P_2=89.6$~d and $K_2=0.44$ ms$^{-1}$ (signal subindex were shifted so $P_1$<$P_2$, and so on), with $\ln\hat{\mathcal{Z}}=-1193.3\pm2.4$, a difference of 24.9 against $1S$.

An extra sinusoid brought multiple solutions in $3S$ at $P_3=655$, 1\,018, and 1\,426~d, the former being the solution with the highest maximum likelihood. The number of live-points was gradually increased until this solution was obtained at least three times out of five consecutive runs. At 5\,000 live-points this solution was obtained consistently (5 out of 5 times), bringing an evidence difference of 24.2 with the $2S$ model.

Changing the sinusoids for Keplerians in the $3K$ model seems to eliminate this problem, since $P_3=655$~d appeared consistently amongst runs (3\,000 live-points). For consistency, we also did the runs with increasing live-points, but when doing this the algorithm appears to get stuck in lower likelihood peaks more often. With 6\,000 live-points we consistently got $P_3=1\,378$~d, with an evidence 15 lower than $P_3=655$~d, and a maximum likelihood 20 lower. The increase in evidence compared to the $3S$ model (3.1) is tinier than both the reference value and the one obtained with the SAR method.

The $4K$ model did not converge to a single solution, obtaining 85.5 d, 111 d, 1011 d, or 1420~d as the fourth signal. The best evidence was obtained by the period $P_4=85.54$~d, while the best likelihood by $P_4=111.23$~d. 

By comparing the differences between each model's best solution, our dyn-rs results match the reference only for $\Delta\mathcal{Z}(H_0, S1)$ and $\Delta\mathcal{Z}(S1, S2)$, with diverging results in increasing models by 5.3, 2.6, and -4.2, respectively. This could be explained by the high standard deviation (e.g., for the $4K$, 4.6 in the reference and 3.8 in dyn-rs) or to sampler setup differences.

On the other hand, doing the same exercise against the SAR run provides much closer differences to the reference: 0.3,  2.2, -1.5,  1.6, and $-0.2$, with the highest values corresponding to $\Delta\mathcal{Z}(2S, 3S)$ and $\Delta\mathcal{Z}(3S, 3K)$, and well within variance.

The \reddemc~runs used 16 temperatures, 256 walkers, and 5\,000 sweeps, increasing by 2\,500 extra sweeps for each model, with 17\,500 sweeps for the $4K$, to facilitate the adaptation stationarity without `hand-tuning' the adaptive parameters. 

\begin{figure}
    \centering
    \includegraphics[width=0.85\columnwidth]{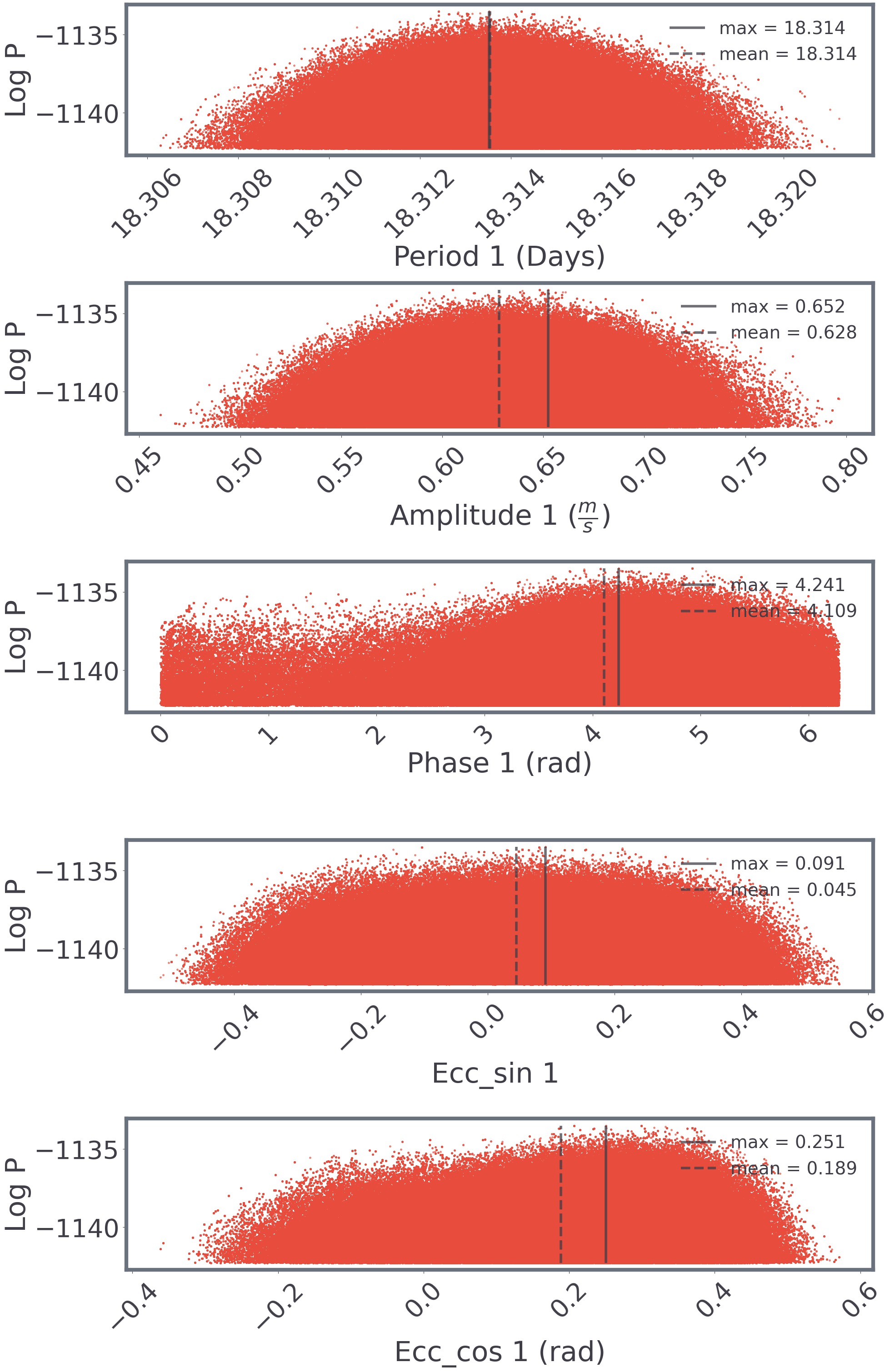}
    \caption{\label{fig:hd20794_posts}HD\,20794 posteriors for the first Keplerian in the $3K$ SAR run.}
\end{figure}

\begin{figure}
\centering
    \includegraphics[width=0.95\columnwidth]{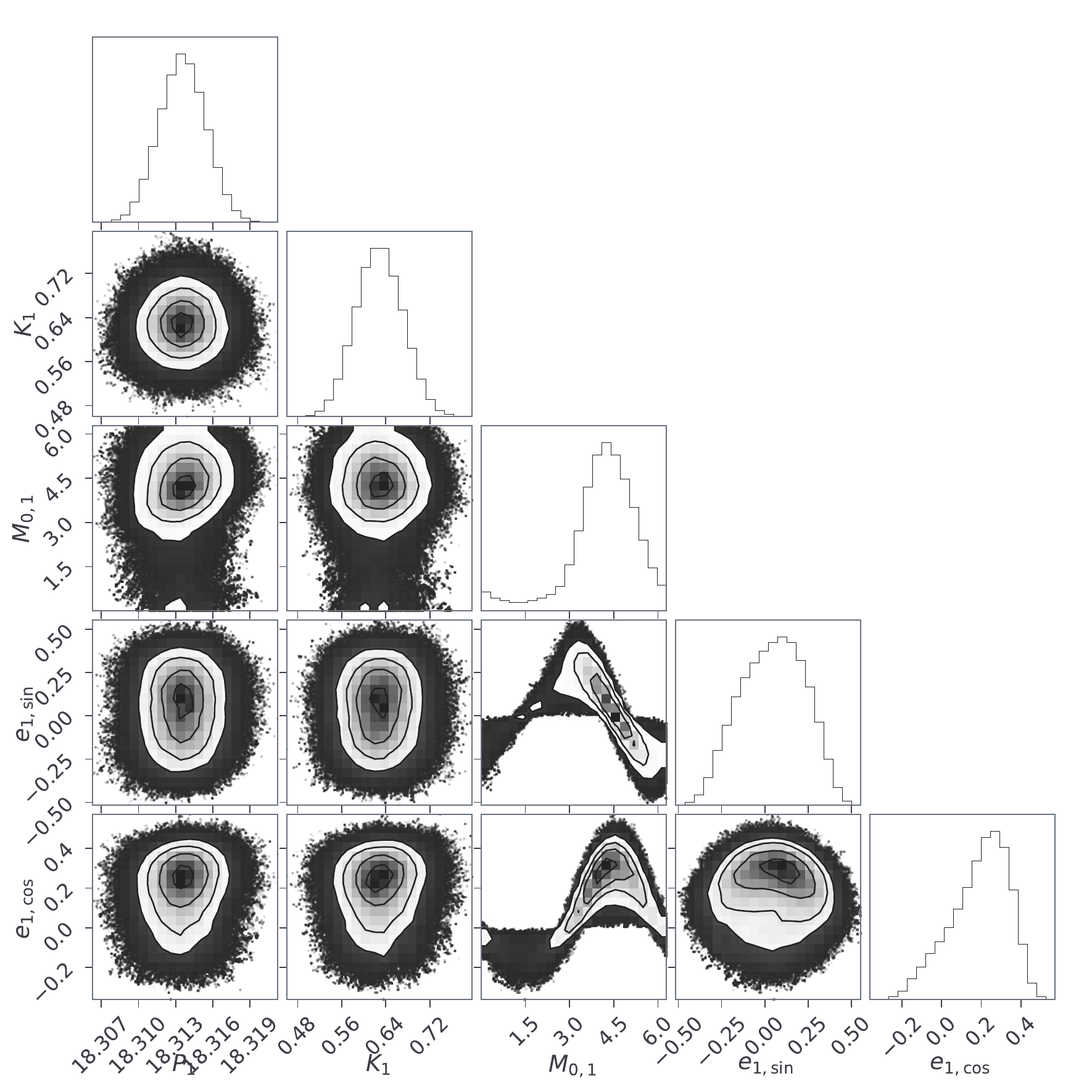}\label{fig:hd20794_corner}

    \includegraphics[width=0.95\columnwidth]{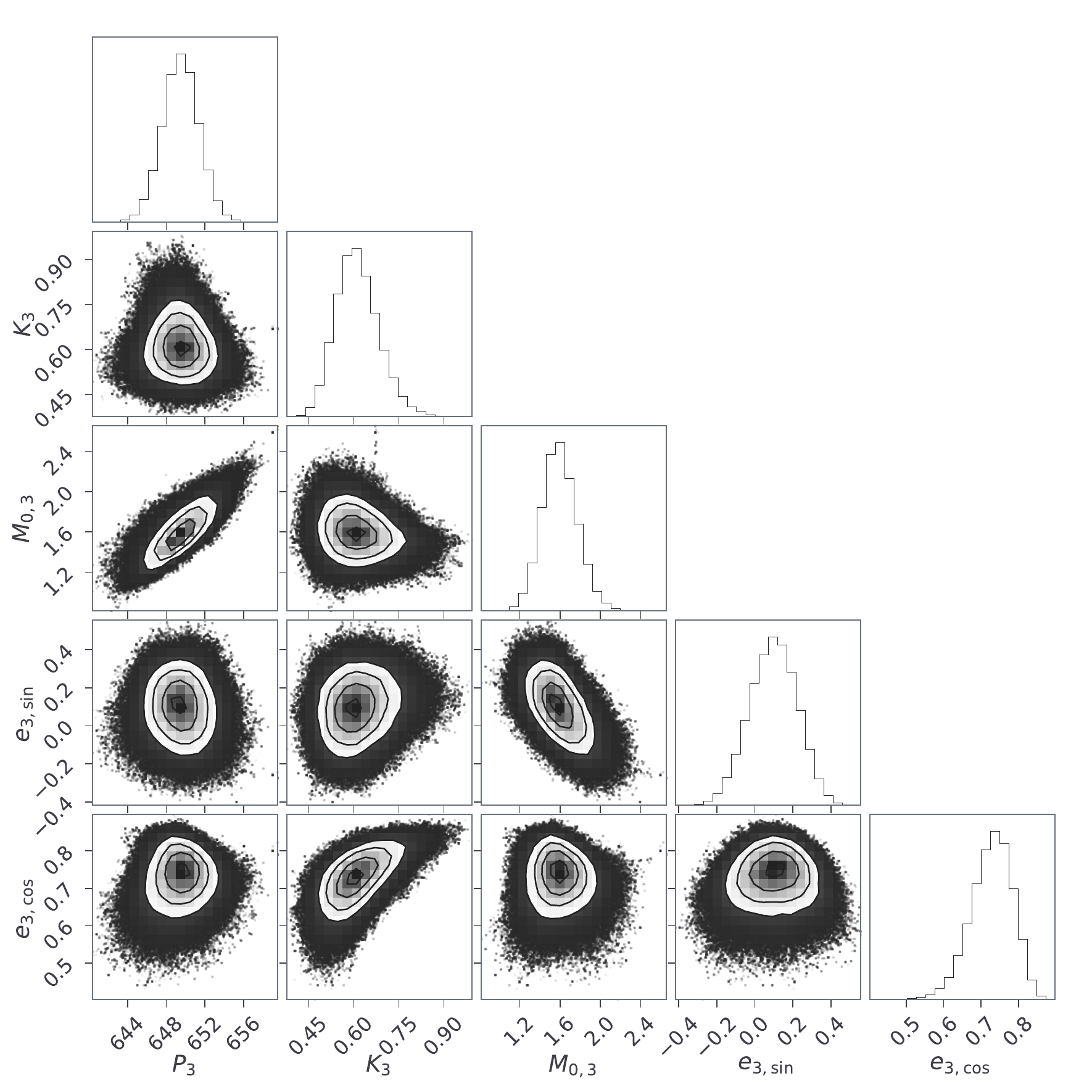}
    \caption{\label{fig:hd20794_corner2}HD\,20794 corner plot for the first (top) and third (bottom) signals in the $3K$ model. Figures made with the \texttt{corner} package \citep{corner}.}
\end{figure}

The $1S$ model consistently found the $P=18.314$~d signal, with an evidence of $\ln\hat{\mathcal{Z}}=-1217.1\pm0.3$, and a difference of 47.9 with the $H_0$ model. Adding a signal found $P_2=89.66$~d consistently, with an evidence improvement of 23.5. The third sinusoid shifted this period slightly to 89.7~d, whilst adding one at $P_3=652.2$~d, and improving the evidence by 29.1. Changing the sinusoids to Keplerians brought an improvement of 5.2 evidence-wise, with eccentricities of $e_1=0.057$, $e_2=0.026$, $e_3=0.467$. The top of the posterior samples for the first Keplerian can be seen in \reffig{fig:hd20794_posts}, with the last two parameters the aforementioned $e_s$ and $e_c$ in \refeq{eq:cv-hou}. Additionally, \reffig{fig:hd20794_corner2} (top) shows a corner plot (scatter plot matrix for multidimensional samples) for the same Keplerian, revealing parameter covariances: $P_1$ and $K_1$ have tight posteriors, with each off-diagonal panel being nearly circular. Both have a little correlation with eccentricity (denoted by their elongation over $e_c, e_s$). The centre of the $e_c, e_s$ panel indicates a low value for eccentricity. The direction of its isotropic shape defines $\omega$. The phase of periastron passage $M_0$ is coupled to the eccentricity vector, the textbook RV degeneracy. The third signal, with a higher well-defined eccentricity, does not present this degeneracy, as seen in \reffig{fig:hd20794_corner2}. 

Finally, the $4K$ model brought both $P_4$$\sim$1440~d and $\sim$111~d, with similar evidences and likelihoods. 

The step decrease in kenits for the more complex model is attributed to the low walker count. It is also worth noting the difference in both consistency and wall-time in these models.  For $3K$ and $4K$ the standard deviation in dyn-rs was 2.6 and 3.8 respectively, compared to 0.4 and 0.9 in \reddemc. Furthermore, the average wall-time in dyn-rs was 150.26±32.46, and 112.77±5.2 minutes, whereas in \reddemc~it was 57.68±2.57, and 77.21±2.42 minutes. Time-wise, the runs were not only significantly shorter, but also more consistent.

\section{Evidence estimators}\label{sec:appendix_math}

\subsection{Geometric-bridge stepping stones} \label{app:ss}

For the temperature ladder $1 = \beta_1 >...>\beta_B \geq 0$, at each $\beta_i$ we have an ensemble of $W$ walkers evolving over $T$ sweeps. For each adjacent pair $(\beta_i, \beta_{i+1})$ and for every sweep, we build a symmetric bridge \citep{meng96_bridge_sampling, gronau17_bridge_sampling} between the two ensembles
\begin{equation} \label{eq:symmetric_bridge}
    \begin{aligned}
        &\text{from $\beta_i$:}\qquad &A_{t, i} &= \frac{1}{W}\sum_{w=1}^{W}\ln\mathcal{L}^{\Delta\beta_i/2}\,, \\
        &\text{and from $\beta_{i+1}$:}\qquad &C_{t, i} &= \frac{1}{W}\sum_{w=1}^{W}\ln\mathcal{L}^{-\Delta\beta_i/2}\,.
    \end{aligned}
\end{equation}

If the ladder is still adapting, we use the per-sweep $\Delta\beta_{i,t}$, otherwise (as in this manuscript) a single averaged $\Delta\beta_{i}$. Then we average over sweeps to get $\mu_{A,i}=\overline{A_{t, i}}$ and $\mu_{C,i}=\overline{C_{t, i}}$. Each ratio $\frac{\mu_{A,i}}{\mu_{C,i}}$ estimates $r_i$ via the geometric bridge. Multiplying all these ratios, or adding their logs, gives the log-evidence estimate
\begin{equation}
    \ln\widehat{\mathcal{Z}}_{\mathrm{SS+}} =\sum_{i=1}^{B-1}\left( \ln\mu_{A,i} - \ln\mu_{C,i} \right)\,.
\end{equation}

For the sampling error $\widehat{\sigma}_S^2$ we treat the per-sweep vectors $\left[ A_{t,\cdot} \mid C_{t,\cdot}\right]$ as a correlated time series across sweeps, estimating the long-run covariance with multivariate OBM, and use the delta method for $g(\mu)=\sum_i \left( \ln\mu_{A,i} - \ln\mu_{C,i} \right)$ to get $\mathrm{Var}[\hat{z}]$ = $\widehat{\sigma}_S^2$ \citep{flegal08_bm_sde, wang_2017_obm}. This method effectively steps along the ladder with a geometric bridge at each step, propagating an autocorrelation-aware uncertainty.

\subsection{Piecewise interpolated thermodynamic integration} \label{app:ti}

For each sweep $t=1,\dots,T$, we take the per-temperature mean log-likelihood $\bar{\ell}_{i, t} = \frac{1}{W}\sum_{w=1}^{W}\ln\mathcal{L}^{\beta_i, t}$. Now, for each sweep we interpolate the curve $\mathbb{E}_{\beta}[U](\beta)$ with PCHIP \citep{pchip}, and integrate, resulting in a series $\hat{z}_t$. The time-average of this series is the evidence estimate
\begin{equation}
    \ln\widehat{\mathcal{Z}}_{\mathrm{TI+}} = \sum_{t=1}^T\hat{z}_t\,.
\end{equation}

To estimate $\widehat{\sigma}_D$ we form a coarser ladder by dropping every other $\beta$, recompute the same PCHIP integral to get $\hat{z}_{t}^{(2)}$, time-average to $\ln\widehat{\mathcal{Z}}_{\mathrm{TI+}}^{(2)}$, and set
\begin{equation}
    \widehat{\sigma}_D = \ln\widehat{\mathcal{Z}}_{\mathrm{TI+}}^{(2)}- \ln\widehat{\mathcal{Z}}_{\mathrm{TI+}}\,.
\end{equation}

To estimate $\widehat{\sigma}_S$, if the ladder is no longer adapting, we treat $\hat{z}_{t}$ as a correlated univariate time-series over the sweeps, estimate its long-run covariance $\Sigma$ with OBM, and set $\widehat{\sigma}_S^2=\frac{\Sigma}{T}$. Otherwise, with a ladder still adapting (under diminishing adaptation), with $N_{\beta}$ the length of the interpolated values, we take the per-sweep trapezoid TI block as
\begin{equation}
    \hat{z}_{t} = \sum_{i=1}^{N_{\beta}} \Delta\beta_{i, t}\cdot S_{i, t}\,, \qquad S_{i, t} = \frac{(\bar{\ell}_{i, t} + \bar{\ell}_{i+1, t})}{2}\,.
\end{equation}

With the $S_{t}$ series, we estimate the multivariate long-run covariance $\Sigma_S$ with multivariate OBM, and propagate with mean widths $\overline{d\beta}=\Delta\beta_t$:
\begin{equation}
    \widehat{\sigma}^2_D = \mathrm{Var}[\hat{z}_{t}] = \overline{d\beta}^{\intercal} \Sigma_S \overline{d\beta}\,.
\end{equation}

Note that for the total error, since $\widehat{\sigma}_D$ and $\widehat{\sigma}_S$ are added in quadrature, they are being treated as independent.
For the adaptative case, we need a stationary adaptation settled in the ladder, which should happen under diminishing adaptations after some sweeps have passed, depending on the hyper-parameters on $\kappa(t)$ (see \refeq{eq:kappa}).

\end{appendix}

\end{document}